\documentclass[aps,preprint,showpacs,showkeys]{revtex4}
\newcommand{\sect}[1]{\setcounter{equation}{0}\section{#1}}
\renewcommand{\theequation}{\arabic{section}.\arabic{equation}}
\newcommand{\app}{\setcounter{section}{0}
\setcounter{equation}{0} \renewcommand{\thesection}{APPENDIX
\Alph{section}}\renewcommand{\theequation}{\Alph{section}.\arabic{equation}}}
\newcommand{\bfm}[1]{\mbox{\boldmath${#1}$}}
\begin{document}
\title{Canonical quantization of nonlinear many body systems}
\author{A.M. Scarfone}
\email{antonio.scarfone@polito.it} \affiliation{Istituto Nazionale
di Fisica della Materia (INFM) and\\ Dipartimento di Fisica
Politecnico di Torino, Corso Duca degli Abruzzi 24, I-10129
Torino, Italy}
\date{\today}

\begin {abstract}
We study the quantization of a classical system
of interacting particles obeying a recently
proposed kinetic interaction principle (KIP) [G.
Kaniadakis, Physica A {\bf 296}, 405 (2001)]. The
KIP fixes the expression of the Fokker-Planck
equation describing the kinetic evolution of the
system and imposes the form of its entropy. In
the framework of canonical quantization, we
introduce a class of nonlinear Schr\"odinger
equations (NSEs) with complex nonlinearities,
describing, in the mean field approximation, a
system of collectively interacting particles
whose underlying kinetics is governed by the KIP.
We derive the Ehrenfest relations and discuss the
main constants of motion arising in this model.
By means of a nonlinear gauge transformation of
third kind it is shown that in the case of
constant diffusion and linear drift the class of
NSEs obeying the KIP is gauge-equivalent to
another class of NSEs containing purely real
nonlinearities depending only on the field
$\rho=|\psi|^2$.
\end {abstract}
 \maketitle
\sect{Introduction}

A wide class of diffusive processes in nature, known as
normal diffusion, are successfully described by the linear
Fokker-Planck equation. Its relation to Boltzmann-Gibbs
entropy (BG-entropy) in the framework of the irreversible
thermodynamics
is well established \cite{deGroot,Prigogine,Glansdorff}.\\
However, nonlinear Fokker-Planck equations
(NFPEs)
\cite{Frank1,Frank2,Chavanis1,Chavanis4,Chavanis2}
and their connection in the field of the
generalized thermodynamics
\cite{Abe,Kaniadakis0,Kaniadakis00} is nowadays
an intense research area. In particular, many
physical phenomena, in presence of memory
effects, nonlocal effects, long-range effects or,
more in general, nonlinear effects, are well
understood with the help of NFPEs.\\ To cite a
few, let us recall the problem of diffusion in
polymers \cite{Ott}, on liquid surfaces
\cite{Bychuk}, in L\'{e}vy flights \cite{Solomon}
and enhanced diffusion in active intracellular
transport \cite {Caspi}. Many anomalous diffusion
systems have a quantum nature, like for instance
charge transport in anomalous solids \cite{Sher},
subrecoil laser cooling \cite{Bardou} and the
aging effect in quantum dissipative systems
\cite{Mauger}.

A still open question concerns the dynamics underlying the
nonlinear kinetics governing the above anomalous systems.
Langevin-like, Fokker-Planck-like or Boltzmann-like
equations have been used by different authors to generate
nonlinear terms in the Schr\"odinger equation with the aim
of describing, in the mean field
approximation, the many quantum particle interactions \cite{Kostin,Schuch1,Schuch,Kaniadakis}.\\
It is now widely recognized that the presence of a
nonlinear drift term as well as the presence of a diffusive
term in a quantum particle current originates complex
nonlinearities in the evolution equation for the $\psi$-wave function.\\
Different examples are known in literature of nonlinear
Schr\"odinger equations (NSEs) originating from the study
of the kinetics governing the many-body quantum system. For
instance, the Doebner-Goldin family equations
\cite{Doebner1} have been introduced from topological
considerations as the most general class of Schr\"odinger
equations compatible with the linear Fokker-Planck
equation. In Ref. \cite{Scarfone3} the authors introduced a
NSE starting from a generalized exclusion-inclusion
principle (EIP) in order to describe systems of quantum
particles with different statistics interpolating with
continuity between the Bose-Einstein and the Fermi-Dirac
ones. In Ref. \cite{Scarfone7}, in the stochastic
quantization framework, starting from the most general
nonlinear kinetics containing a nonlinear drift term and
compatible with a linear diffusion term, a class of NSEs
with a complex nonlinearity was obtained.

Recently, a kinetic interaction principle (KIP) has been
proposed \cite{Kaniadakis1} to define a special collective
interaction among the $N$-identical particles of a
classical system. On the one hand, the KIP imposes the form
of the generalized entropy associated with the system,
while on the other hand it governs the evolution of the
system toward equilibrium by fixing the expression of the
nonlinear current of particles
in the NFPE, thus governing the kinetics underlying the system.\\
The link between the generalized entropic functional and
the corresponding NFPE can also be obtained starting from a
maximum entropic production principle. In Refs.
\cite{Chavanis1,Chavanis4}, taking into account a
variational principle maximizing the dissipation rate of a
generalized free energy, the authors obtained a NFPE in the
Smoluchowski limit. The same NFPE was obtained in Ref.
\cite{Chavanis2} from a stochastic process described by a
generalized Langevin equation where the strength of the
noise is assumed to depend on the density of the particle.

In the present paper we perform the quantization of a
classical system obeying KIP, where
the statistical information is supplied by a very general entropy.\\
Up to today, different methods have been proposed for the
microscopic description of systems. Schr\"odinger's wave
mechanics, Heisenberg's matrix mechanics or Feynman's
path-integral mechanics are some of the many. Another
approach is given by the hydrodynamic theory of quantum
mechanics originally owing to Madelung \cite{Madelung} and
de Broglie \cite{Broglie} and successively reconsidered by
Bohm \cite{Bohm} in connection with his theory of hidden
variables.\\
In the hydrodynamic formulation of quantum mechanics, the
complex linear Schr\"odinger equation is replaced by two
real nonlinear differential equations for two independent
fields: the probability density and its velocity field.
Basically, such equations are formally similar to the
equations of continuity and the Euler
equation of ordinary hydrodynamics.\\
This formalism is fruitful, as in the present situation,
when the expression of the quantum continuity equation is
inherited from the one describing the kinetics of the
ancestor classical system. However, for a complete quantum
mechanical description, besides the continuity equation, we
need to know if and how we should generalize the Euler
equation that describes the dynamics of the system. In this
paper, in order to fix the nonlinear terms in the Euler
equation, we require that the whole
model be formulated in the canonical formalism.\\
We obtain a class of NSEs with complex nonlinearity
describing a quantum system of interacting particles
obeying the KIP in the mean field approximation. We study
the case of a quantum system undergoing a constant
diffusion process. The generalization to the case of a
nonconstant diffusive process is also presented at the end
of the paper. It is shown that the form of the entropy of
the ancestor classical system fixes the nonlinearity
appearing in the evolution equation. By means of a recently
proposed nonlinear gauge transformation
\cite{Doebner1,Scarfone1,Scarfone4} this family of
evolution equations is transformed into another one
describing a nondiffusive process. In particular, when the
kinetics of the system is governed by a linear drift term,
the new family of NSEs contains a purely real nonlinearity
depending only on
the density of particles $\rho=|\psi|^2$.\\
As working examples we present the quantization of some
classical systems described by entropies already known in
the literature: BG-entropy, Tsallis-entropy \cite{Tsallis},
Kaniadakis-entropy \cite{Kaniadakis1} and the interpolating
quantum statistics entropy \cite{Quarati1}.

The plan of the paper is the following. In Section II we
recall the relation between a given generalized entropy and
the associated NFPE describing the kinetic evolution of the
classical system in the nonequilibrium thermodynamic
framework. This kinetic equation is justified on the ground
of KIP. In Section III, firstly first present an overall
summing up of the hydrodynamic formulation of the linear
Schr\"odinger equation, then we generalize the method to
quantize the classical system obeying EIP. The Hamiltonian
formulation of this model is presented and a family of NSEs
with complex nonlinearity is obtained. In Section IV we
study the Ehrenfest relations and discuss the conserved
mean quantities. In Section V, the nonlinear gauge
transformation is introduced. Some relevant examples are
presented in Section VI. The final Section VII present
comments and conclusions. In Appendix A we give the
derivation of the Ehrenfest relations while in Appendix B
we briefly discuss the generalization of the model for a
quantum system whose kinetics undergoes a nonconstant
diffusive process.


\sect{Nonlinear Fokker-Planck equation}

Our starting point, according to nonlinear kinetics, is to
relate the production of the entropy of a classical system
to a Fokker-Planck equation. This can be accomplished by
following the classical approach to diffusion
\cite{deGroot,Prigogine}.\\ We start by assuming a very
general trace-form expression for the entropy (throughout
this paper, we use units with the Boltzmann constant
$k_{_{\rm B}}$ set equal to unity)
\begin{equation}
S(\rho)=-\int d{\bfm x}\int d\rho\,\ln\kappa(\rho) \
,\label{entropy}
\end{equation}
where $\kappa(\rho)$ is an arbitrary functional of the density
particles field $\rho=\rho(t,\,{\bfm x})$, with ${\bfm
x}\equiv(x_{_1},\,\cdots,\,x_{_n})$
a point in the $n$-dimensional space.\\
The constraints
\begin{equation}
\int \rho\,d{\bfm x}=1 \ ,\label{norm}
\end{equation}
on the normalization and
\begin{equation}
\int{\mathcal E}({\bfm x})\,\rho\,d{\bfm x}=E \ ,\label{energy}
\end{equation}
total energy of the system, with ${\mathcal E}({\bfm
x})={\bfm p}^2/2\,m+V({\bfm x})$ the energy for each
particle, are accounted for by introducing the constrained
entropic functional
\begin{equation}
{\mathcal S(\rho)}=-\int d{\bfm x}\int
d\rho\,\ln\kappa(\rho)-\beta\int {\mathcal E}({\bfm
x})\,\rho\,d{\bfm x}-\beta^\prime\int \rho\,d{\bfm x} \
.\label{funentropy}
\end{equation}
The two constants $\beta$ and $\beta^\prime$ are the Lagrange
multipliers associated with constraints (\ref{norm}) and (\ref{energy}).\\
Quite generally, the evolution of the field $\rho$ in the
configuration space is governed by the continuity equation
\begin{equation}
\frac{\partial\,\rho}{\partial\,t}+{\bfm \nabla}\cdot{\bfm J}=0 \
,\label{cont}
\end{equation}
with
${\bfm\nabla}\equiv(\partial/\partial\,x_{_1},\,\cdots,\,\partial/\partial\,x_{_n})$,
and assures the conservation of the constraint (\ref{norm}) in
time.\\
We assume a nonlinear relation between the current $\bfm J$ and
the constrained thermodynamic force
\begin{equation}
\bfm{\mathcal F}(\rho)={\bfm\nabla}\left(\frac{\delta{\mathcal
S}}{\delta\rho}\right) \ ,\label{force}
\end{equation}
by posing
\begin{equation}
{\bfm J}=D\,\gamma(\rho)\,\bfm{\mathcal F}(\rho) \ ,\label{d}
\end{equation}
with $D$ the diffusion coefficient and $\gamma(\rho)$ still an
arbitrary functional of $\rho$.\\ Putting Eq. (\ref{d}) in Eq.
(\ref{cont}), and taking into account the expression of $\mathcal
S$ given in Eq. (\ref{funentropy}) we obtain the following
continuity equation
\begin{equation}
\frac{\partial\rho}{\partial t}+{\bfm
\nabla}\cdot\Big\{-D\,\gamma(\rho){\bfm \nabla}\Big[\beta\,
{\mathcal E}({\bfm x})+\beta^\prime+\ln\kappa(\rho)\Big]\Big\}=0 \
.\label{cont1}
\end{equation}
Introducing drift velocity
\begin{equation}
{\bfm u}_{\rm drift}=-D\,\beta\,{\bfm\nabla}\,{\mathcal E}({\bfm
x}) \ ,
\end{equation}
Eq. (\ref{cont1}) takes the form of a NFPE for the field $\rho$
\begin{equation}
\frac{\partial\,\rho}{\partial\,t}+{\bfm \nabla}\cdot\Big[{\bfm
u}_{\rm drift}\,\gamma(\rho)-D\,f(\rho)\,{\bfm\nabla}\,\rho\Big]=0
\ ,\label{cont2}
\end{equation}
where
\begin{equation}
f(\rho)=\gamma(\rho)\,
\frac{\partial\,\ln\,\kappa(\rho)}{\partial\,\rho} \ .
\end{equation}
Total current ${\bfm J}={\bfm J}_{\rm drift}+{\bfm J}_{\rm
diff}$ is the sum of a nonlinear drift current ${\bfm
J}_{\rm drift}={\bfm u}_{\rm drift}\,\gamma(\rho)$, and a
nonlinear diffusion current ${\bfm J}_{\rm
diff}=-D\,f(\rho)\,{\bfm \nabla}\,\rho$, different from
Fick's standard one ${\bfm J}_{\rm
Fick}=-D\,{\bfm\nabla}\,\rho$, which is recovered by posing
$\gamma(\rho)=\kappa(\rho)=\rho$.\\
Eq. (\ref{cont2}) describes a class of nonlinear diffusive
processes
varying the functionals $\gamma(\rho)$ and $\kappa(\rho)$.\\
We observe that for any given entropy (\ref{entropy}) an infinity
of associated NFPEs exists, one for any choice of $\gamma(\rho)$.

In Refs. \cite{Chavanis1,Chavanis4}, starting from a
variational principle which maximizes the dissipation rate
of a generalized free energy functional, substantially
equivalent to Eq. (\ref{funentropy}), a NFPE in the
position space as in Eq. (\ref{cont2}) has been obtained.
The same NFPE (\ref{cont2}) was also obtained in Ref.
\cite{Chavanis2}, starting from a stochastic process
described by a generalized Langevin equation, where the
strength of the noise is assumed to depend on the density
of the particle. The nonlinear current, as in Eq.
(\ref{d}), is given by the gradient of the functional
derivative of a generalized free
energy equivalent to Eq. (\ref{funentropy}).\\
In Ref. \cite{Frank1} the problem of the NFPE derived from
generalized linear nonequilibrium thermodynamics was also
discussed at length.

At equilibrium, the particle current must vanish, and from
Eq. (\ref{force}) it follows
\begin{equation}
\ln\kappa(\rho_{\rm eq})+\beta\,{\mathcal E}({\bfm
x})+\beta^\prime=0 \ ,\label{equi}
\end{equation}
where, without loss of generality, we posed the integration
constant equal to zero (otherwise it can be included in the
Lagrange multiply $\beta^\prime$).\\ We obtain the
equilibrium distribution of the system
\begin{equation}
\rho_{\rm eq}=\kappa^{-1}\Big(\exp\left(-\beta\,{\mathcal E}({\bfm
x})-\beta^\prime\right)\Big) \ .\label{distry}
\end{equation}
In particular, with the choice $\kappa(\rho)=e\,\rho$, Eq.
(\ref{entropy}) reduces to standard BG-entropy and Eq.
(\ref{distry}) gives the
well-known Gibbs-distribution.\\

Let us now justify Eq. (\ref{cont2}) starting from the
kinetic approach introduced in \cite{Kaniadakis1} through
the KIP. In accordance with the arguments presented in Ref.
\cite{Kaniadakis1}, we consider the following classical
Markovian process
\begin{equation}
\frac{\partial\,\rho}{\partial\,t}=\int\left[\pi(t,{\bfm
y}\rightarrow{\bfm x})-\pi(t,{\bfm x}\rightarrow{\bfm
y})\right]\,d{\bfm y} \ ,\label{ee}
\end{equation}
describing the kinetics of a system of $N$-identical
interacting particles.\\ For transition probability
$\pi(t,{\bfm x}\rightarrow{\bfm y})$ we assume a suitable
expression in terms of the populations of the initial site
$\bfm x$ and
the final site $\bfm y$.\\
According to KIP we pose
\begin{equation}
\pi(t,{\bfm x}\rightarrow{\bfm y})=r(t,\,{\bfm x},\,{\bfm x}-{\bfm
y}) \,\gamma(\rho,\,\rho^\prime) \ ,\label{tran}
\end{equation}
where $\rho\equiv\rho(t,\,{\bfm x})$ and
$\rho^\prime\equiv\rho(t,\,{\bfm y})$ are the particle
density functions in the starting site $\bfm x$ and in the
arrival site $\bfm y$ respectively, whereas $r(t,\,{\bfm
x},\,{\bfm x}-{\bfm y})$ is the transition rate which
depends only on the starting $\bfm x$ and arrival $\bfm y$
sites, during particle transition ${\bfm x}\rightarrow{\bfm
y}$.\\ The functional $\gamma(\rho,\,\rho^\prime)$ can be
factorized in
\begin{equation}
\gamma(\rho,\,\rho^\prime)=a(\rho)\,b(\rho^\prime)\,c(\rho,\,\rho^\prime)
\ .\label{gamma}
\end{equation}
The first factor $a(\rho)$ is a functional of the particle
population $\rho$ of the starting site and satisfies the
boundary condition $a(0)=0$, since if the starting site is
empty transition probability is equal to zero. The second
factor $b(\rho^\prime)$ is a functional of the particle
population $\rho^\prime$ at the arrival site, and satisfies
the condition $b(0)=1$, because the transition probability
does not depend on the arrival site if particles are absent
there. Finally, the third factor $c(\rho,\,\rho^\prime)$
takes into account that the populations of
the two sites can eventually affect the transition collectively and symmetrically.\\
The expression of the functional $b(\rho^\prime)$ plays a
very important role in the particle kinetics because it can
stimulate or inhibit the transition ${\bfm
x}\rightarrow{\bfm y}$, allowing, in this way, interactions
originating from collective effects.\\ With the assumptions
made in Eqs. (\ref{tran}) and (\ref{gamma}) for transition
probability, according to the Kramers-Moyal expansion and
assuming the first neighbor approximation, we can expand up
to the second order the quantities $r(t,\,{\bfm y},\,{\bfm
y}-{\bfm x})\,\gamma(\rho(t,\,{\bfm y}),\,\rho(t,\,{\bfm
x}))$ and $\gamma(\rho(t,\,{\bfm x}),\,\rho(t,\,{\bfm y}))$
in Taylor series of ${\bfm y}={\bfm x}+{\bfm u}$ and ${\bfm
y}={\bfm x}-{\bfm u}$, respectively, in an interval around
$\bfm x$, for ${\bfm u}\ll{\bfm x}$. We obtain
\begin{eqnarray}
\nonumber& &r(t,\,{\bfm x}+{\bfm u},\,{\bfm
u})\,\gamma\Big(\rho(t,\,{\bfm x}+{\bfm u}),\,\rho(t,\,{\bfm
x})\Big)\\ \nonumber&&\hspace{-5mm}=\left\{r(t,\,{\bfm y},\,{\bfm
u})\,\gamma\Big(\rho(t,\,{\bfm y}),\,\rho(t,\,{\bfm
x})\Big)+\frac{\partial}{\partial\,y_{_i}}\left[r(t,\,{\bfm
y},\,{\bfm u})\,\gamma\Big(\rho(t,\,{\bfm y}),\,\rho(t,\,{\bfm
x})\Big)\right]\,u_{_i} \right.\\
&&+\left.{1\over2}\,\frac{\partial^2}{\partial\,y_{_i}\,\partial\,y_{_j}}
\left[r(t,\,{\bfm y},\,{\bfm u})\,\gamma\Big(\rho(t,\,{\bfm
y}),\,\rho(t,\,{\bfm
x})\Big)\right]\,u_{_i}\,u_{_j}\right\}_{{\bfm y}\rightarrow{\bfm
x}} \ ,\label{sv1}
\end{eqnarray}
and
\begin{eqnarray}
\nonumber\hspace{-8mm}\gamma\Big(\rho(t,\,{\bfm
x}),\,\rho(t,\,{\bfm x}-{\bfm u})\Big)&
&=\Bigg\{\gamma\Big(\rho(t,\,{\bfm x}),\,\rho(t,\,{\bfm
y})\Big)-\frac{\partial}{\partial\,y_{_i}}\gamma\Big(\rho(t,\,{\bfm
x}),\,\rho(t,\,{\bfm
y})\Big)\,u_{_i}\Bigg.\\
&
&\left.+{1\over2}\,\frac{\partial^2}{\partial\,y_{_i}\,\partial\,y_{_j}}
\,\gamma\Big(\rho(t,\,{\bfm x}),\,\rho(t,\,{\bfm
y})\Big)\,u_{_i}\,u_{_j}\right\}_{{\bfm y}\rightarrow{\bfm x}} \
.\label{sv2}
\end{eqnarray}
Using Eqs. (\ref{sv1}) and (\ref{sv2}) in Eq. (\ref{tran}),
from Eq. (\ref{ee}) it follows
\begin{equation}
\frac{\partial\,\rho}{\partial\,t}=\frac{\partial}{\partial
x_{_i}}\left[\left(
\zeta_{_i}+\frac{\partial\,\zeta_{_{ij}}}{\partial
x_{_j}}\right)\,\gamma(\rho)+\zeta_{_{ij}}\,\gamma(\rho)\,
\frac{\partial}{\partial\,x_{_j}}\,\ln\kappa(\rho)\right] \
,\label{kip}
\end{equation}
with $i=1,\,\cdots,\,n$ and summation over repeated indices is
assumed.\\ In Eq. (\ref{kip})
\begin{equation}
\gamma(\rho)\equiv\gamma(\rho,\,\rho^\prime)\Bigg|_{\rho=\rho^\prime}
\ ,
\end{equation}
and
\begin{equation}
\kappa(\rho)=\frac{a(\rho)}{b(\rho)} \ ,\label{lambda}
\end{equation}
while the coefficients $\zeta_{_i}$ and $\zeta_{_{ij}}$ are
given by
\begin{equation}
\zeta_{_i}=\int r(t,\,{\bfm y},\,{\bfm u})\,u_{_i}\,d{\bfm u} \
,\label{eej}
\end{equation}
\begin{equation}
\zeta_{_{ij}}=\frac{1}{2}\int r(t,\,{\bfm y},\,{\bfm
u})\,u_{_i}\,u_{_j}\,d{\bfm u} \ .\label{ees}
\end{equation}
Defining $(u_{_i})_{\rm
drift}=-\zeta_{_i}-\partial\,\zeta_{_{ij}}/\partial\,x_{_j}$,
the $i$-th component of ${\bfm u}_{\rm drift}$, and
assuming the independence of motion in different directions
of the isotropic configuration space we can pose
$\zeta_{_{ij}}=D\,\delta_{_{ij}}$. It is easy to see that
Eq. (\ref{kip}) reduces to Eq. (\ref{cont2}).

In conclusion we observe that Eq. (\ref{cont2}) is a NFPE
in the Smoluchowski limit since it describes a kinetic
process in the position space rather than in the phase
space. This is a suitable form for the quantum treatment of
the following sections. The passage from the NFPE in the
phase space to the NFPE in the position space was
rigorously elaborated in Ref. \cite{Chavanis3} in the limit
of strong friction, by means of a Chapman-Enskog-like
expansion.


\sect{Canonical quantization}

\subsection{Quantization in the hydrodynamic
representation}\label{hydrodynamic} In the hydrodynamic
representation, the quantum mechanics formulation, can
readily be obtained from the standard Schr\"odinger
equation
\begin{equation}
i\,\hbar\,\frac{\partial\,\psi}{\partial\,t}=-\frac{\hbar^2}{2\,m}\,\Delta\,\psi+V({\bfm
x})\, \psi \ ,\label{sch}
\end{equation}
where $V({\bfm x})$ is a real external potential. The
complex field $\psi\equiv\psi(t,\,{\bfm x})$ describing the
quantum system is related to the hydrodynamic fields
$\rho(t,\,{\bfm x})$ and $\Sigma(t,\,{\bfm x})$ through
polar decomposition \cite{Bohm,Madelung}
\begin{equation}
\psi(t,\,{\bfm x})=\rho^{1/2}(t,\,{\bfm
x})\,\exp\left(\frac{i}{\hbar}\,\Sigma(t,\,{\bfm x})\right) \
.\label{polar}
\end{equation}
Eq. (\ref{sch}) is separated into a couple of real
equations
\begin{eqnarray}
&&m\,\frac{\partial\,\widehat{\bfm
v}}{\partial\,t}+m\,\left(\widehat{\bfm
v}\cdot{\bfm\nabla}\right)\,\widehat{\bfm
v}={\bfm\nabla}\,\left(\frac{\hbar^2}{2\,m}\,\frac{\Delta\,\sqrt{\rho}}{\sqrt{\rho}}
-V({\bfm x})\right) \ ,\label{h1}\\
&&\frac{\partial\,\rho}{\partial\,t}+{\bfm\nabla}\cdot{\bfm
j}_{_0}=0 \ ,\label{h2}
\end{eqnarray}
where quantum velocity $\widehat{\bfm v}$, which in the
linear case coincides with quantum drift velocity
$\widehat{\bfm u}_{\rm drift}$, is related to the phase
$\Sigma(t,\,{\bfm x})$ through
\begin{equation}
m\,\widehat{\bfm v}={\bfm\nabla}\,\Sigma(t,\,{\bfm x}) \
,\label{vquant}
\end{equation}
and
\begin{equation}
{\bfm j}_{_0}=\rho\,\widehat{\bfm v} \ ,\label{jc}
\end{equation}
is the same relationship between current and velocity of the
standard hydrodynamic theory. We remark that the quantum current
(\ref{jc}) contains only a linear drift term.\\
According to the orthodox interpretation of quantum
mechanics the quantity $\rho(t,\,{\bfm x})=|\psi(t,\,{\bfm
x})|^2$ represents the position probability density of the
system normalized as $\int\rho(t,\,{\bfm x})\,d{\bfm x}=1$.

Eqs. (\ref{h1})-(\ref{jc}) form the basis of the
hydrodynamic formulation which consists of a quasi
classical approach to quantum mechanics. In this picture
the evolution of the system can be interpreted in terms of
a flowing fluid with density $\rho(t,\,{\bfm x})$
associated with a local velocity field $\widehat{\bfm
v}(t,\,{\bfm x})$. The dynamics of such fluid is described
by the Euler equation (\ref{h1}) and is governed by forces
arising not only from the external field ${\bfm F}_{\rm
ext}({\bfm x})=-{\bfm\nabla}\,V({\bfm x})$ but also from an
additional potential
$U_q=-(\hbar^2/2\,m)\,\Delta\sqrt\rho/\sqrt\rho$ known as
the quantum potential \cite{Bohm}. Remarkably, the
expectation value for the quantum force vanishes at all
times, i.e. $\langle-{\bfm\nabla}\,U_q\rangle=0$. Finally,
the continuity equation (\ref{h2}) assures the conservation
of the normalization
of wave function $\psi$ during the evolution of the system.\\
Let us remark that the quantum fluid has a very special property.
Because $\Sigma(t,\,{\bfm x})$ is a potential field for the
quantum velocity, the quantum fluid is irrotational. As a
consequence, in the linear Schr\"odinger theory, a non vanishing
vorticity $\bfm \omega$, defined by
\begin{equation}
{\bfm\omega}={\bfm\nabla}\times\widehat{\bfm v} \
,\label{vorticity}
\end{equation}
is possible only at the nodal region where neither
$\Sigma(t,\,{\bfm x})$ nor ${\bfm\nabla}\,\Sigma(t,\,{\bfm
x})$ are well defined. At such a point
${\bfm\nabla}\times{\bfm\nabla}\,\Sigma(t,\,{\bfm x})$ does
not vanish in general, thus leading to the appearance of
point-like vortices.

Finally, putting Eq. (\ref{vquant}) into Eq. (\ref{h1}) we obtain
\begin{equation}
\frac{\partial\,\Sigma}{\partial\,t}+\frac{\left({\bfm\nabla}\,\Sigma\right)^2}{2\,m}
-\frac{\hbar^2}{2\,m}\,\frac{\Delta\sqrt{\rho}}{\sqrt{\rho}}+V({\bfm
x})=0 \ .\label{si1}
\end{equation}
This equation, in the classical limit $\hbar\to0$, reduces to the
Hamilton-Jacobi equation for the function
$\Sigma$.\\
Eqs. (\ref{h2}) and Eq. (\ref{si1}) can be obtained by means of
the Hamiltonian equations
\begin{eqnarray}
&&\frac{\partial\,\Sigma}{\partial\,t}=-\frac{\delta\,
H}{\delta\,\rho} \ ,\label{rhos1}\\
&&\frac{\partial\,\rho}{\partial\,t}=\frac{\delta\,
H}{\delta\,\Sigma} \ ,\label{rhos2}
\end{eqnarray}
where the Hamiltonian
\begin{equation}
H=\int{\cal H}(\rho,\,\Sigma)\,d{\bfm x} \ ,\label{ha}
\end{equation}
with
\begin{equation}
{\mathcal
H}(\rho,\,\Sigma)=\frac{({\bfm\nabla}\,\Sigma)^2}{2\,m}\,\rho+\frac{\hbar^2}{8\,m}
\,\frac{\left({\bfm\nabla}\,\rho\right)^2}{\rho}+V({\bfm x})\,\rho
\ .\label{ham11}
\end{equation}
represents the total energy of the quantum system.

\subsection{The many-body quantum system}

Let us now generalize the method described above by
replacing the linear continuity equation Eq. (\ref{h2})
with the more general one obtained in analogy with the
continuity equation (\ref{cont2}) describing the kinetics
of a classical system obeying KIP. In the following we
assume that the quantum system undergoes a constant
diffusion process with $D=const.$

We begin by introducing the wave function
$\psi\equiv\psi(t,\,{\bfm x})$ describing, in the mean field
approximation, a system of quantum interacting particles. We
postulate that the following NSE describes the evolution equation
of the system
\begin{equation}
i\,\hbar\,\frac{\partial\,\psi}{\partial\,t}=
-\frac{\hbar^2}{2\,m}\,\Delta\,\psi+\Lambda(\psi^\ast,\,\psi)\,\psi+V({\bfm
x})\,\psi \ ,\label{schroedinger1}
\end{equation}
where $\Lambda(\psi^\ast,\,\psi)=W(\psi^\ast,\,\psi)+i\,{\mathcal
W}(\psi^\ast,\,\psi)$ is a complex nonlinearity, with
$W(\psi^\ast,\,\psi)$ and ${\mathcal W}(\psi^\ast,\,\psi)$ the
real and the imaginary part, respectively.\\
Using polar decomposition (\ref{polar}), Eq.
(\ref{schroedinger1}) is separated into a couple of real
nonlinear evolution equations for phase and amplitude
\begin{eqnarray}
&&\frac{\partial\,\Sigma}{\partial\,t}+\frac{\left({\bfm\nabla}\,\Sigma\right)^2}{2\,m}
+U_q+W(\rho,\,\Sigma)+V({\bfm
x})=0 \ ,\label{s}\\
&&\frac{\partial\,\rho}{\partial\,t}+{\bfm\nabla}\cdot{\bfm
j}_{_0}-\frac{2}{\hbar}\,\rho\,{\mathcal W}(\rho,\,\Sigma)=0 \
.\label{rho}
\end{eqnarray}
We require that both Eqs. (\ref{s}) and (\ref{rho}) can be
obtained through the Hamilton equations
(\ref{rhos1})-(\ref{rhos2}) and, to accommodate
nonlinearities $W(\rho,\,\Sigma)$ and ${\cal
W}(\rho,\,\Sigma)$, we introduce in the Hamiltonian density
${\cal H}$ an additional real nonlinear potential
$U(\rho,\,\Sigma)$ which describes the collective
interaction between the particles belonging to the system
\begin{equation}
{\mathcal
H}(\rho,\,\Sigma)=\frac{({\bfm\nabla}\,\Sigma)^2}{2\,m}\,\rho+\frac{\hbar^2}{8\,m}
\,\frac{\left({\bfm\nabla}\,\rho\right)^2}{\rho}+U(\rho,\,\Sigma)+V({\bfm
x})\,\rho \ .\label{ham1}
\end{equation}
By means of Eqs. (\ref{rhos1}) and (\ref{rhos2}) it follows that
the nonlinear functionals $W(\rho,\,\Sigma)$ and ${\mathcal
W}(\rho,\,\Sigma)$ are related to the nonlinear potential
$U(\rho,\,\Sigma)$ as
\begin{eqnarray}
&&W(\rho,\,\Sigma)=\frac{\delta\, }{\delta\,\rho}\int
U(\rho,\,\Sigma)\,d{\bfm x} \ ,\label{c}\\ &&{\mathcal
W}(\rho,\,\Sigma)=\frac{\hbar}{2\,\rho}\,\frac{\delta}{\delta\,\Sigma}\int
U(\rho,\,\Sigma)\,d{\bfm x} \ .\label{cw}
\end{eqnarray}
We assume that the quantum fluid satisfies a continuity
equation formally equal to the classical one described by
the NFPE (\ref{cont2}). By matching Eq. (\ref{rho}) with
Eq. (\ref{cont2}) we obtain the expression ${\cal W}$ and,
accounting for Eq. (\ref{cw}), we have the nonlinear
potential $U(\rho,\,\Sigma)$. Finally, the nonlinearity
$W(\rho,\,\Sigma)$, which follows from Eq. (\ref{c}),
together with the quantum potential $U_q$ and the external
potential $V({\bfm x})$, describes the dynamic behavior of
the quantum fluid according to Eq. (\ref{s}).

We observe that if the following equation holds
\begin{equation}
\frac{\delta}{\delta\,\Sigma}\,\int U(\rho,\,\Sigma)\,d{\bfm
x}={\bfm\nabla}\cdot{\bfm F}(\rho,\,\Sigma) \ ,\label{uf}
\end{equation}
with ${\bfm F}(\rho,\,\Sigma)$ an arbitrary functional, taking
into account Eq. (\ref{cw}), Eq. (\ref{rho}) becomes
\begin{equation}
\frac{\partial\,\rho}{\partial\,t}+{\bfm\nabla}\cdot\left[{\bfm
j}_{_0}-{\bfm F}(\rho,\,\Sigma)\right]=0 \ .\label{rho1}
\end{equation}
Eq. (\ref{uf}) is fulfilled if functional
$U(\rho,\,\Sigma)$
depends on phase $\Sigma$ only through its spatial derivatives \cite{Scarfone1}.\\
Introducing the quantum drift velocity
\begin{equation}
\widehat{\bfm u}_{\rm drift}=\frac{{\bfm\nabla}\,\Sigma}{m} \ ,
\end{equation}
which in the linear case coincides with the quantum
velocity $\widehat{\bfm v}$ given in Eq. (\ref{vquant}),
and by comparing Eq. (\ref{rho1}) with Eq. (\ref{cont2}) we
have
\begin{equation}
{\bfm
F}(\rho,\,\Sigma)=\frac{{\bfm\nabla}\,\Sigma}{m}\,\Big[\rho-\gamma(\rho)\Big]+D\,f(\rho)\,
\,{\bfm\nabla}\,\rho \ .
\end{equation}
By integrating Eq. (\ref{cw}), the nonlinear potential assumes the
expression
\begin{equation}
U(\rho,\,\Sigma)=\frac{({\bfm\nabla}\,\Sigma)^2}{2\,m}\,\Big[\gamma(\rho)-\rho\Big]
-D\,f(\rho)\,{\bfm\nabla}\rho\cdot{\bfm\nabla \Sigma}+{\widetilde
U}(\rho) \ ,\label{u}
\end{equation}
where $\widetilde U(\rho)$ is an arbitrary real potential
depending only on field $\rho$. Eqs. (\ref{rhos1}) and
(\ref{rhos2}) give the following coupled nonlinear
evolution equations
\begin{eqnarray}
\nonumber&&\frac{\partial\,\Sigma}{\partial\,t}+
\frac{({\bfm\nabla\,\Sigma})^2}{2\,m}\,\frac{\partial\,\gamma(\rho)}{\partial\,\rho}
-\frac{\hbar^2}{2\,m}\, \frac{\Delta\,\sqrt{\rho}}{\sqrt{\rho}}
+m\,D\,f(\rho)\,{\bfm\nabla}\cdot\left(\frac{{\bfm
j}_{_0}}{\rho}\right)+G(\rho)+V({\bfm x})=0 \ ,\label{s1}\\
\end{eqnarray}
\begin{eqnarray}
&&\frac{\partial\,\rho}{\partial\,t}+{\bfm\nabla}\cdot
\left[\frac{{\bfm \nabla}\,\Sigma}{m}\,\gamma(\rho)-D\,f(\rho)
\,{\bfm\nabla}\,\rho\right]=0 \ ,\label{r1}
\end{eqnarray}
where $G(\rho)=\delta\,\int{\widetilde U}(\rho)\,d{\bfm
x}/\delta\,\rho$.\\
In the classical limit $\hbar\to0$ Eq. (\ref{s1}) becomes a
nonlinear Hamilton-Jacobi equation for function $\Sigma$.
It differs from the classical one owing to the presence of
the nonlinear term which functionally depends on both
$\rho$ and $\Sigma$. We recall that such a nonlinearity was
introduced consistently
with the requirement of a final canonical formulation of the theory.\\
We stress once again that in the approach described in this
paper, we start from a nonlinear generalization of the
continuity equation that gives us only information on the
kinetics. This equation is not enough to completely
determine the time evolution of the quantum system. As a
consequence, we have ample freedom in the definition of
nonlinear potential $U(\rho,\,\Sigma)$. Such freedom is
reflected in the arbitrary functional $\widetilde U(\rho)$
which cannot be fixed only on the basis of the kinetic
equation. There are many possible dynamic behaviors, one
for any choice of $\widetilde U(\rho)$, compatible with the
same kinetics. The nonlinear potential ${\widetilde
U}(\rho)$ can be used to describe other possible
interactions among the many particles of the system that
have an origin different from the one introduced by the
kinetic equation (\ref{r1}).

Actually, Eq. (\ref{r1}) is a quantum continuity equation
for field $\rho$ with a nonlinear quantum current given by
\begin{equation}
{\bfm j}=\frac{{\bfm
\nabla}\,\Sigma}{m}\,\gamma(\rho)-D\,f(\rho)\,{\bfm\nabla}\,\rho \
.\label{ncurrent}
\end{equation}
We observe that, differently from the hydrodynamic
formulation of the linear quantum mechanics, where the
Bohm-Madelung fluid is irrotational, in nonlinear quantum
theory the situation can be very different. In fact, by
defining quantum velocity through Eq. (\ref{jc}), from Eq.
(\ref{ncurrent}) we have
\begin{equation}
m\,\widehat{\bfm
v}={\gamma(\rho)\over\rho}\,{\bfm\nabla}\,\Big[\Sigma-m\,D\,\ln\kappa(\rho)\Big]
\ ,\label{vq1}
\end{equation}
which states the relationship between quantum velocity
$\widehat{\bfm v}$ and quantum drift velocity
$\widehat{\bfm u}_{\rm drift}$ for the nonlinear
case.\\Expression (\ref{vq1}) can be justified in terms of
Clebsh potentials. In fact, as is well known, a
nonvanishing vorticity can be accounted for in the
Schr\"odinger theory by introducing three potentials $\mu$,
$\nu$ and $\lambda$ related to quantum velocity through the
relation
\begin{equation}
m\,\widehat{\bfm v}={\bfm\nabla}\,\mu+\nu\,{\bfm\nabla}\,\lambda \
.\label{vq}
\end{equation}
Vorticity ${\bfm\omega}$ assumes a nonvanishing expression
given by
\begin{equation}
{\bfm\omega}={1\over
m}{\bfm\nabla}\,\nu\times{\bfm\nabla}\,\lambda \
.\label{vorticity1}
\end{equation}
By comparing Eq. (\ref{vq}) with Eq. (\ref{vq1}) we readily obtain
$\mu=const$, $\nu=\gamma(\rho)/\rho$ and
$\lambda=\Sigma-m\,D\,\ln\kappa(\rho)$, respectively, and Eq.
(\ref{vorticity1}) becomes
\begin{equation}
{\bfm\omega}={1\over
m}{\bfm\nabla}\,\left(\frac{\gamma(\rho)}{\rho}\right)\times{\bfm\nabla}\,
\Sigma \ ,
\end{equation}
with no any contribution from the diffusive term. The
irrotational case is recovered in linear drift
$\gamma(\rho)=\rho$.

The final expression of the NSE (\ref{schroedinger1}) is given by
\begin{equation}
i\,\hbar\,\frac{\partial\,\psi}{\partial\,t}=
-\frac{\hbar^2}{2\,m}\,\Delta\,\psi+\Big[W(\rho,\,\Sigma)+i\,{\mathcal
W}(\rho,\,\Sigma)\Big]\,\psi+V({\bfm x})\,\psi \
,\label{schroedinger2}
\end{equation}
with the nonlinearities
\begin{eqnarray}
W(\rho,\,\Sigma)={m\over2}\,
\left(\frac{\partial\,\gamma(\rho)}{\partial\,\rho}-1\right)\,\left(\frac{{\bfm
j}_0}{\rho}\right)^2+m\,D\,f(\rho)\,{\bfm\nabla}\cdot\left(\frac{{\bfm
j}_{_0}}{\rho}\right)+G(\rho) \ , \label{www}
\end{eqnarray}
and
\begin{eqnarray}
\hspace{-3mm}{\mathcal
W}(\rho,\,\Sigma)=-\frac{\hbar}{2\,\rho}\,{\bfm\nabla}\,\Bigg\{[\gamma(\rho)-\rho]
\left(\frac{{\bfm j}_0}{\rho}\right)\Bigg\}
+\frac{\hbar\,D}{2\,\rho}\,{\bfm\nabla}\cdot\left[f(\rho )
\,{\bfm\nabla}\,\rho\right] \ . \label{cwww}
\end{eqnarray}
Eqs. (\ref{www})-(\ref{cwww}) differ from the one obtained
in Ref. \cite{Scarfone7} where a family of NSE was derived
in the stochastic quantization framework starting from the
most general nonlinear classical kinetics compatible with
constant diffusion coefficient $D=\hbar/2\,m$. In
particular, the real nonlinearity $W$ arising in the
stochastic quantization is found to depend only on field
$\rho$, in contrast with expression (\ref{www}), where
functional $W$ depends on
both fields $\rho$ and $\Sigma$.\\
Remarkably, we observe that when the kinetics of the system
is governed by a linear drift, with $\gamma(\rho)=\rho$,
the expression of nonlinear terms (\ref{www}) and
(\ref{cwww}) simplify to
\begin{equation}
W(\rho,\,\Sigma)=m\,D\,\widetilde
f(\rho)\,{\bfm\nabla}\cdot\left(\frac{{\bfm
j}_{_0}}{\rho}\right)+G(\rho) \ ,
\end{equation}
and
\begin{equation}
{\mathcal W}(\rho,\,\Sigma)=
\frac{\hbar\,D}{2\,\rho}\,{\bfm\nabla}\cdot\left[\widetilde
f(\rho)\,\ln\kappa(\rho) \,{\bfm\nabla}\,\rho\right] \ ,
\end{equation}
where $\widetilde f(\rho)=\rho\,(\partial/\partial\,\rho)\,\ln\kappa(\rho)$.\\
They are determined only through functional $\kappa(\rho)$
which also defines the entropy (\ref{entropy}) of the
ancestor classical system.


\sect{Ehrenfest relations and conserved
quantities}\label{Ehrenfest}

In this section we study the time evolution of the most
important physical observables of the system described by
the Hamiltonian density (\ref{ham1}) with the nonlinear
potential (\ref{u}): mass center, linear and angular
momentum and total energy.
The proofs are given in Appendix A.\\
Let us recall that, given an Hermitian operator ${\cal
O}={\cal O}^\dag$ associated with a physical observable,
its time evolution is given by
\begin{equation}
\frac{d}{d\,t}\langle{\cal O}\,\rangle={i\over\hbar}\int\left(
\frac{\delta\,H}{\delta\,\psi}\,{\cal O}\,\psi-\psi^\ast\,{\cal
O}\,\frac{\delta\,H}{\delta\,\psi^\ast}\right)\,d{\bfm
x}+\Bigg\langle\frac{\partial\,{\cal O}}{\partial\,t}\Bigg\rangle
\ ,\label{ehrenfest}
\end{equation}
where the mean value $\langle{\cal O}\,\rangle=\int\psi^\ast{\cal
O}\,\psi\,d{\bfm x}$. The last term in Eq. (\ref{ehrenfest}) takes
into account a possible explicit time dependence on the operator
$\cal O$.\\
Observing that the NSE (\ref{schroedinger2}) can be written in
\begin{equation}
i\,\hbar\,\frac{\partial\,\psi}{\partial\,t}=\textsf{H}\,\psi \
,\label{HHH}
\end{equation}
where
\begin{equation}
\textsf{H}=
-\frac{\hbar^2}{2\,m}\,\Delta+W(\rho,\,\Sigma)+i\,{\cal
W}(\rho,\,\Sigma)+V({\bfm x}) \ ,\label{HHH1}
\end{equation}
Eq. (\ref{ehrenfest}) assumes the equivalent expression
\begin{equation}
\frac{d}{d\,t}\langle{\cal
O}\,\rangle={i\over\hbar}\,\Big\langle\Big[{\rm
Re}\,\textsf{H},\,{\cal
O}\Big]\Big\rangle+{1\over\hbar}\,\Big\langle\Big\{{\rm
Im}\,\textsf{H},\,{\cal O}\Big\}\Big\rangle
+\Bigg\langle\frac{\partial\,{\cal O}}{\partial\,t}\Bigg\rangle \
,\label{ehrenfestnew}
\end{equation}
where $[\cdot,\,\cdot]$ and $\{\cdot,\,\cdot\}$ indicate the
commutator and the anticommutator, respectively.

By choosing ${\cal O}={\bfm x}$, from Eq. (\ref{ehrenfest}) we
obtain the first Ehrenfest relation for the time evolution of the
mass center of the system
\begin{equation}
{\bfm v}_{\rm mc}\equiv\frac{d}{d\,t}\,\langle{\bfm
x}\rangle=\Bigg\langle \frac{\gamma(\rho)}{\rho}\,\widehat{\bfm
u}_{\rm drift}\Bigg\rangle \ .\label{prima}
\end{equation}
We observe that only drift nonlinearity appears in this
equation whereas the diffusion term makes no contribution.
Eq. (\ref{prima}) states that, quite generally, ${\bfm
v}_{\rm mc}$ is not a motion constant. This fact implies
that the quantum system is not Galilei invariant. The
origin of the nonconservation of ${\bfm v}_{\rm cm}$ can be
found in the difference between quantity ${\bfm p}_{\rm
mc}=m\,{\bfm v}_{\rm mc}$ and the expectation value of the
momentum operator ${\bfm
p}\equiv\langle-i\,\hbar\,{\bfm\nabla}\rangle=\int\rho\,{\bfm\nabla}\,\Sigma\,d{\bfm
x}$. These two quantities are equivalent only in the linear
drift case. Differently from the former, the latter is in
all cases conserved during the time evolution of the
system, in absence of the external potential. This can be
shown by means of the second Ehrenfest relation which
follows from Eq. (\ref{ehrenfest}) by posing ${\cal
O}=-i\,\hbar\,{\bfm\nabla}$
\begin{equation}
\frac{d}{d\,t}\,\langle{\bfm p}\rangle=\Big\langle{\bfm F}_{\rm
ext}({\bfm x})\Big\rangle \ .\label{seconda}
\end{equation}
The time evolution of the expectation value of momentum is
governed only by external potential $V({\bfm x})$. On the
average, the KIP introduce no effect on the dynamics of the
system. This is a consequence of the invariance of
nonlinearity $W[\rho,\,\Sigma]+i\,{\cal W}[\rho,\,\Sigma]$
under uniform space translation.

In the same way, accounting for the invariance of
nonlinearity for uniform rotations, the third Ehrenfest
relation follows
\begin{equation}
\frac{d}{d\,t}\,\langle{\bfm L}\rangle=\Big\langle{\bfm M}_{\rm
ext}({\bfm x})\Big\rangle \ ,\label{terza}
\end{equation}
where ${\bfm M}_{\rm ext}({\bfm x})={\bfm x}\times{\bfm
F}_{\rm ext}({\bfm x})$ is the momentum of the external
force field. Eq. (\ref{terza}) is obtained from Eq.
(\ref{ehrenfest}) after posing ${\cal O}={\bfm
x}\times(-i\,\hbar\,{\bfm \nabla})$. Again, the nonlinear
terms introduced by KIP as well as nonlinearity $G(\rho)$
make no contribution, on the average, to angular momentum.

Finally, the last relation concerns the total energy of the system
given by the Hamiltonian $E\equiv H$. By posing
\begin{equation}
{\cal
O}=-\frac{\hbar^2}{2\,m}\,\Delta+{1\over\rho}\,U(\rho,\,\Sigma)+V({\bfm
x}) \ ,
\end{equation}
we have $\langle{\cal O}\rangle\equiv E$ and from Eq.
(\ref{ehrenfest}) we obtain
\begin{equation}
\frac{d\,E}{d\,t}=0 \ .\label{quarta}
\end{equation}

In conclusion, for a constant diffusion process we have
shown that in absence of the external potential the system
admits three constants of motion: total linear momentum
$\langle{\bfm p}\rangle$, total angular momentum
$\langle{\bfm L}\rangle$ and total energy $E$. Such
conserved quantities, according to the Noether theorem,
follow as a consequence of the invariance of the system
under uniform space-time translation and uniform rotation.
Moreover, the system is also invariant for global U(1)
transformation which implies conservation of the
normalization of field $\psi$ throughout the evolution of
the system.\\ In Appendix B we briefly discuss the case of
a quantum system with a diffusion coefficient $D(t,\,{\bfm
x})$ that depends on time and position. This space-time
dependence destroys the invariance of the system under
uniform space-time translation and space rotation. As a
consequence, all quantities $\langle{\bf p}\rangle$,
$\langle{\bf L}\rangle$ and $E$ are no longer conserved,
even for a vanishing external potential.

It should be remarked that the results discussed here,
although very general in that they are independent of the
form of nonlinearities $W$ and $\cal W$, are valid only for
the class of the canonical systems. In literature there are
many noncanonical NSEs, obtained starting from certain
physically motivated conditions, which are worthy of being
taken into account. For these equations, the expression of
$\texttt{H}$ appearing on the right hand side of the
Schr\"odinger equation cannot be obtained from Eqs.
(\ref{rhos1}) and (\ref{rhos2}) by means of a Hamiltonian
function $H=\int{\cal H}\,d{\bfm x}$.\\
Despite this, even for these noncanonical systems the time
evolution of the mean values of the quantum operators
associated with the observables can be derived through Eq.
(\ref{ehrenfestnew}), but what is important is that these
operators can assume a different definition with respect to
the one given in the canonical theory. For instance, in the
canonical framework the energy is supplied by the
Hamiltonian $H$ of the system, whereas in a noncanonical
theory it is identified with the operator
$i\,\hbar\,\partial /
\partial\,t\equiv\texttt{H}$. (We remark that in the canonical framework
$H$ and $\texttt{H}$ are, in general, different
quantities). Moreover, for a noncanonical theory,
conservation of the energy and the momentum do not follow
merely from the principle of invariance of the system under
space-time translation. Their time evolution depends on the
expression of the nonlinearities appearing in the
Schr\"odinger equation. All of this clearly causes a
profound
difference in the resulting Ehrenfest relations.\\
For instance, in Ref. \cite{Schuch1} a noncanonical
Schr\"odinger equation with complex nonlinearity was
derived starting from a Fokker-Planck equation for density
field $\rho$ by assuming some physically justified
separability conditions. The resulting evolution equation
has the real and the imaginary nonlinearity given by
$W(\rho,\,\Sigma)=\gamma\,(\Sigma-\langle\Sigma\rangle)$
and ${\cal
W}(\rho,\,\Sigma)=(\hbar\,D/2)\,\Delta\rho/\rho$,
respectively, where $\gamma$ is a constant related to
diffusion coefficient $D$ and such that $D\to0$ if
$\gamma\to0$. It is easy to see that such nonlinearities
cannot be obtained starting from a nonlinear potential
$U(\rho,\,\Sigma)$ through Eqs. (\ref{c}) and (\ref{cw}).
The system described by this NSE turns out to be dumped and
dissipative, even in presence of a constant diffusive
process. In fact, it can be shown that, following Ref.
\cite {Schuch1}, from Eq. (\ref{ehrenfestnew}) it follows
$d\,\langle{\bfm p}\rangle/d\,t=\langle{\bfm F}_{\rm
ext}\rangle-\gamma\,\langle{\bfm p}\rangle$ and
$d\,E/d\,t\equiv d\,\langle
\texttt{H}\rangle/d\,t=-(\gamma/m)\,\langle{\bfm
p}^2\rangle$, which is a very different situation with
respect to the one discussed in the
present paper, with the exception of the trivial case $\gamma=0$.\\

\sect{Gauge equivalence}\label{Gauge equivalence}

We introduce a nonlinear gauge transformation of the third
kind \cite{Scarfone1}
\begin{equation}
\psi\rightarrow\phi=\psi\,\exp\left(-\frac{i}{\hbar}\,m\,D\,\ln\kappa(\rho)\right)
\ ,\label{gauge}
\end{equation}
which, being a unitary transformation, does not change the
amplitude of wave function $|\psi|^2=|\phi|^2=\rho$, and
transforms the phase $\Sigma$ of the old field $\psi$, into
phase $\sigma$ of the new field $\phi$ according to the
equation
\begin{equation}
\sigma=\Sigma-m\,D\,\ln\kappa(\rho) \ .\label{phase}
\end{equation}
Consequently, the nonlinear current (\ref{ncurrent}) takes the
expression
\begin{equation}
{\bfm j}\to\widetilde{\bfm
j}=\frac{{\bfm\nabla}\sigma}{m}\,\gamma(\rho) \ .\label{jt}
\end{equation}
with only a nonlinear drift term.\\
Let us observe that, at the classical level, the similar
transformation
\begin{equation}
{\bfm u}_{\rm drift}^\prime={\bfm
u}_{\rm drift}-D\,{\bfm\nabla}\,\ln\kappa(\rho) \ ,\label{trtr}
\end{equation}
changes total current ${\bfm J}\rightarrow{\bfm
J}^\prime={\bfm u}_{\rm drift}^\prime\,\gamma(\rho)$ into
another one consisting only of a nonlinear drift term.

Performing the transformation (\ref{gauge}), Eq.
(\ref{schroedinger2}) becomes
\begin{equation}
i\,\hbar\,\frac{\partial\,\phi}{\partial\,t}=
-\frac{\hbar^2}{2\,m}\,\Delta\,\phi+ \left[\widetilde
W(\rho,\,\sigma)+i\,\widetilde{\mathcal W}(\rho,\,\sigma)\right]
\,\phi+V({\bfm x})\,\phi \ ,\label{schroedinger31}
\end{equation}
where the new nonlinearities $\widetilde W(\rho,\,\sigma)$ and
$\widetilde{\mathcal W}(\rho,\,\sigma)$ are given by
\begin{eqnarray}
\nonumber\widetilde W(\rho,\,\sigma)={m\over2}\,
\left(\frac{\partial\,\gamma(\rho)}{\partial\,\rho}-1\right)\,\left(\frac{\widetilde{\bfm
j}_0}{\rho}\right)^2+m\,D^2\,\Bigg[f_1(\rho)\,\Delta\,\rho+f_2(\rho)\,
\left({\bfm\nabla}\,\rho\right)^2 \Bigg]+G(\rho) \ ,\\ \label{w}
\end{eqnarray}
with $\widetilde{\bfm j}_0=\rho\,{\bfm\nabla}\,\sigma/m$,
\begin{eqnarray}
&&f_1(\rho)=\gamma(\rho)\,\left[\frac{\partial}{\partial\,\rho}\,\ln\kappa(\rho)\right]^2
\ ,\\
&&f_2(\rho)={1\over2}\,\frac{\partial\,f_{1}(\rho)}{\partial\,\rho}
\ ,
\end{eqnarray}
and
\begin{equation}
\widetilde{\mathcal
W}(\rho,\,\Sigma)=-\frac{\hbar}{2\,\rho}\,{\bfm\nabla}\,\Bigg\{[\gamma(\rho)-\rho]
\left(\frac{\widetilde{\bfm j}_0}{\rho}\right)\Bigg\} \
.\label{gcw}
\end{equation}
Eq. (\ref{schroedinger31}) is still a NSE with a complex
nonlinearity due to the presence of the nonlinear drift term in
the quantum current expression (\ref{jt}).

Basically, both equations (\ref{schroedinger2}) and
(\ref{schroedinger31}) are different NSEs describing the
same physical system. This is a consequence of the unitary
structure of the transformation (\ref{gauge}) which implies
that the probability position density for field $\psi$ and
field $\phi$ assumes the same value at any instant of time
\cite{Doebner1}.

In the case of $\gamma(\rho)=\rho$ expressions (\ref{w})
and (\ref{gcw}) can be simplified and the NSE
(\ref{schroedinger31}) assumes the form
\begin{eqnarray}
\nonumber
i\,\hbar\,\frac{\partial\,\phi}{\partial\,t}=&-&\frac{\hbar^2}{2\,m}\,\Delta\,\phi
+m\,D^2\,\Bigg[\widetilde
f_{1}(\rho)\,\Delta\,\rho+\widetilde f_2(\rho)\,
\left({\bfm\nabla}\,\rho\right)^2\Bigg]\,\phi+G(\rho)\,\phi+V({\bfm
x})\,\phi \ ,\\ \label{schroedinger3}
\end{eqnarray}
with
\begin{eqnarray}
&&\widetilde
f_{1}(\rho)=\rho\,\,\left[\frac{\partial}{\partial\,\rho}\,\ln\kappa(\rho)\right]^2 \ ,\\
&&\widetilde f_2(\rho)={1\over2}\,\frac{\partial\,\widetilde
f_{1}(\rho)}{\partial\,\rho} \ ,
\end{eqnarray}
which contains a purely real nonlinearity depending only on
field $\rho$.\\
We observe that although Eq. (\ref{gauge}) transforms the
nonlinear current into another one without the diffusive
term, NSEs (\ref{schroedinger31}) and (\ref{schroedinger3})
contain a dependence from on diffusion coefficient $D$.\\
The NSE (\ref{schroedinger31}) is still canonical. It can be
obtained from the following Hamiltonian density
\begin{equation}
{\mathcal
H}(\rho,\,\sigma)=\frac{({\bfm\nabla}\,\sigma)^2}{2\,m}\,\rho+\frac{\hbar^2}{8\,m}
\,\frac{\left({\bfm\nabla}\,\rho\right)^2}{\rho}+\widehat
U(\rho,\,\sigma)+V({\bfm x})\,\rho \ ,\label{ht}
\end{equation}
with nonlinear potential
\begin{equation}
\widehat
U(\rho,\,\sigma)=\frac{({\bfm\nabla}\,\sigma)^2}{2\,m}\,\Big[\gamma(\rho)-\rho\Big]-
{m\,D^2\over2}\,f_1(\rho)\,\left({\bfm\nabla}\,\rho\right)^2+\widetilde
U(\rho) \ .\label{ut}
\end{equation}
In this sense Eq. (\ref{gauge}) defines a canonical
transformation.

In conclusion, let us make the following observation. Eq.
(\ref{schroedinger31}) admits the following continuity
equation
\begin{equation}
\frac{\partial\,\rho}{\partial\,t}+{\bfm\nabla}\cdot
\left[\frac{{\bfm\nabla}\sigma}{m}\,\gamma(\rho)\right]=0 \
.\label{cnc}
\end{equation}
A natural question is: what kind of NSE is obtained if we
quantize a classical system obeying the continuity equation
$\partial\,\rho/\partial\,t+{\bfm\nabla}\cdot{\bfm J}^\prime=0$
with the method described above?\\
We easily have
\begin{eqnarray}
\nonumber i\,\hbar\,\frac{\partial\,\phi}{\partial\,t}=
&-&\frac{\hbar^2}{2\,m}\,\Delta\,\phi-{m\over2}\,
\left(\frac{\partial\,\gamma(\rho)}{\partial\,\rho}-1\right)\,\left(\frac{\widetilde{\bfm
j}_0}{\rho}\right)^2\,\phi\\&-&i\,\frac{\hbar}{2\,\rho}\,{\bfm\nabla}\,\Bigg\{[\gamma(\rho)-\rho]
\left(\frac{\widetilde{\bfm j}_0}{\rho}\right)\Bigg\}
\,\phi+G(\rho)\,\phi+V({\bfm x})\,\phi \
,\label{schroedinger311}
\end{eqnarray}
where now $\rho $ and $\sigma$ are independent fields
representing the amplitude and phase of wave function
$\phi$. Eq. (\ref{schroedinger311}) can be derived through
the Hamiltonian density (\ref{ht}) with nonlinear potential
\begin{equation}
\widehat
U_{_1}(\rho,\,\sigma)=\frac{({\bfm\nabla}\,\sigma)^2}{2\,m}\,\Big[\gamma(\rho)-\rho\Big]
+\widetilde U(\rho)\ .\label{utt}
\end{equation}
Potentials (\ref{ut}) and (\ref{utt}) differ for the
quantity
\begin{equation}
\overline U(\rho)=\widehat U(\rho,\,\sigma)-\widehat
U_{_1}(\rho,\,\sigma)=-
{m\,D^2\over2}\,f_1(\rho)\,\left({\bfm\nabla}\,\rho\right)^2 \ ,
\end{equation}
which depends only on field $\rho$. This nonlinear
potential $\overline U(\rho)$ does not affect the
continuity equation and thus cannot be obtained starting
directly from Eq. (\ref{cnc}).


\sect{Some examples}\label{exemples}

To illustrate the relevance and applicability of the theory
described in the previous sections, we derive and discuss
some different NSEs obtained starting from kinetic
equations known in literature. In the following Section,
for simplicity's sake we omit the arbitrary nonlinear
potential $\widetilde U(\rho)$ and focus our attention only
on the effect yield through the potential introduced by the
KIP.

\subsection{Boltzmann-Gibbs-entropy} It is well known
that when the many body system is governed by short-range
interactions, or when interaction energy is neglecting with
respect to the total energy of the system, the suitable
entropic functional is given by the BG-entropy
\begin{equation}
S_{{\rm BG}}(\rho)=-\int\rho\,\ln\left(\rho\right)\,d{\bfm x} \
.\label{SBG}
\end{equation}
This entropy arises from Eq. (\ref{entropy}) by posing
$\kappa(\rho)=e\,\rho$ with $a(\rho)=e\,\rho$ and
$b(\rho)=1$. It is readily seen that
$\gamma(\rho)=e\,\rho\,c(\rho)$.\\ Among the many NFPEs
compatible with entropy (\ref{SBG}) we consider the
simplest case of linear drift by posing $c(\rho)=1/e$. Then
the continuity equation (\ref{r1}) becomes the standard
linear Fokker-Planck equation
\begin{equation}
\frac{\partial \rho}{\partial t}+{\bfm\nabla}\cdot\left({\bfm
j}_{_0}-D\,{\bfm\nabla}\,\rho\right)=0 \ ,\label{Shannon}
\end{equation}
whereas the evolution equation for the quantum system is
given by the following NSE
\begin{eqnarray}
i\,\hbar\,\frac{\partial\,\psi}{\partial\,t}=&-&\frac{\hbar^2}{2\,m}\,\Delta\,\psi+
m\,D\,{\bfm\nabla}\cdot\left(\frac{{\bfm
j}_{_0}}{\rho}\right)\,\psi+i\,\frac{\hbar}{2}\,D\,\frac{\Delta\,\rho}{\rho}\,\psi+V({\bfm
x})\,\psi \ ,\label{DG}
\end{eqnarray}
which is recognized as the canonical sub-family of the
class of Doebner-Goldin equations parameterized by
diffusion coefficient $D$. We recall that Eq.
(\ref{Shannon}) was obtained in the quantum mechanics
theory starting from the study of the physical
interpretation of a certain family of
diffeomorphismin group \cite{Doebner1}.\\
By performing gauge transformation (\ref{gauge}), Eq.
(\ref{DG}) becomes
\begin{equation}
i\,\hbar\,\frac{\partial\,\phi}{\partial\,t}=-\frac{\hbar^2}{2\,m}\,\Delta\,\phi
+m\,D^2\,\left[\frac{\Delta\,\rho}{\rho}
-{1\over2}\left(\frac{{\bfm\nabla}\,\rho}{\rho}\right)^2\right]\,\phi+V({\bfm
x})\,\phi \ ,\label{DG1}
\end{equation}
which was studied previously in \cite{Guerra}. In
particular, Eq. (\ref{DG1}) is equivalent to the following
linear Schr\"odinger equation
\begin{equation}
i\,k^{\!\!\!\!\!-}\,\frac{\partial\,\chi}{\partial\,t}=-\frac{{k^{\!\!\!\!\!-}}^2}
{2\,m}\,\Delta\,\chi+V({\bfm x})\,\chi \ ,\label{lsch}
\end{equation}
with $k^{\!\!\!\!\!-}=\hbar\,\sqrt{1-(2\,m\,D/\hbar)^2}$
and field $\chi$ is related to hydrodynamic fields $\rho$
and $\sigma$ through the relation
$\chi=\rho^{1/2}\,\exp(i\,\sigma/k^{\!\!\!\!\!-})$.\\
This appear to be an interesting result. By quantizing a
classical system described by MB-entropy the standard
linear Schr\"odinger equation was obtained. In this
equation the nonlinear terms describing the interaction
between the many particles of the quantum system are
absent. This is in accordance with the general statement
that MB-entropy is suitable for describing systems with no
(or negligible) interaction among the particles.

\subsection{Generalized entropies}
In presence of long-range interactions or memory effects
persistent in time, it has been argued that MB-entropy may
not be appropriate in describing such systems. For this
reason, many different
versions of Eq. (\ref{SBG}) have been proposed in literature.\\
Very recently, Ref. \cite{Scarfone9,Scarfone8} introduced a
bi-parametric deformation of the logarithmic function
\begin{equation}
\ln_{_{\{\kappa,r\}}}(x)=\frac{x^{r+\kappa}-x^{r-\kappa}}{2\,\kappa}
\ ,\label{krlog}
\end{equation}
which reduces, in the $(\kappa,\,r)\to(0,\,0)$ limit, to the
standard logarithm: $\ln_{_{\{0,0\}}}(x)=\ln x$. By replacing the
logarithmic function in Eq. (\ref{SBG}) with its generalized
version (\ref{krlog}), we obtain a bi-parametric family of
generalized entropies
\begin{equation}
S_{_{\{\kappa,r\}}}(\rho)=-\int
\rho\,\ln_{_{\{\kappa,r\}}}\left(\rho\right)\,d{\bfm x} \
,\label{skr}
\end{equation}
introduced, for the first time, in Refs.
\cite{Mittal,Sharma}. Remarkably, this family of entropies
includes, as special cases, some generalized entropies,
well known in literature, used in the study of systems
exhibiting distribution with asymptotic power law behavior.
Among them we can cite Tsallis-entropy \cite{Tsallis} which
follows by posing $r=\pm|\kappa|$
\begin{equation}
S_{_q}(\rho)=\int\frac{\rho^q-\rho}{1-q}\,d{\bfm x} \
,\label{tsallis}
\end{equation}
with $q=1\pm2\,|\kappa|$ and Kaniadakis-entropy
\cite{Kaniadakis1}, for $r=0$
\begin{equation}
S_{_{\{\kappa\}}}(\rho)=-\int
\frac{\rho^{1+\kappa}-\rho^{1-\kappa}}{2\,\kappa}\,d{\bfm x} \
.\label{sk}
\end{equation}

Both these entropies, as well as other one-parameter
deformed entropies, originated from Eq. (\ref{skr})
\cite{Scarfone8}, can be employed to describe generalized
statistical systems such as, for instance, charge particles
in electric and magnetic fields \cite{Rossani},
2d-turbulence in pure-electron plasma \cite{Boghosian},
Bremsstrahlung \cite{Souza} and anomalous diffusion of the
correlated and L\'{e}vy type
\cite{Borland,Zanette}.\\
In addition to the many applications where Tsallis-entropy
has been employed \cite{Tsallisbiblio}, Kaniadakis-entropy
(\ref{sk}) has been successfully applied in the description
of the energy distribution of fluxes of cosmic rays
\cite{Kaniadakis1}, whereas the entropy in (\ref{skr}) with
$\kappa^2=(r+1)^2-1$ has been applied in the generalized
statistical mechanical study of $q$-deformed oscillators in
the
frame-work of quantum-groups \cite{Abe2}.\\
Despite the topics recalled above, there is currently great
interest in studying quantum systems with long-range
microscopic interactions. Systems such as quantum wires,
which are now possible in practice thanks to recent
technological advances, require on the theoretical ground,
the development of a quantum (nonlinear) theory capable of
capturing the emergent facts \cite{Nazareno}.

The entropy in (\ref{skr}) arises from Eq. (\ref{entropy})
by posing
\begin{equation}
\ln\kappa(\rho)=\lambda\,\ln_{_{\{{\scriptstyle
\kappa,r}\}}}\left(\frac{\rho}{\alpha}\right) \ ,\label{lk}
\end{equation}
with
$\lambda=(1+r-\kappa)^{(r+\kappa)/2\,\kappa}/(1+r+\kappa)^{(r-\kappa)/2\,\kappa}$
and
$\alpha=[(1+r-\kappa)/(1+r+\kappa)]^{1/2\,\kappa}$.\\
Among the many different possibilities, we discuss the case
of linear drift with $\gamma(\rho)=\rho$. By taking into
account Eq. (\ref{lk}) we have continuity equation
(\ref{r1}) with
\begin{equation}
f(\rho)=a_+\rho^{r+\kappa}-a_-\rho^{r-\kappa} \ ,
\end{equation}
where
$a_\pm=(r\pm\kappa)\,(1+r\pm\kappa)/2\,\kappa$ are constants.\\
The associated NSE assumes the expression
\begin{eqnarray}
i\,\hbar\,\frac{\partial\,\phi}{\partial\,t}=&-&\frac{\hbar^2}{2\,m}
\Delta\,\phi+m\,D^2\,{f(\rho)\over\rho}\,
\left[f(\rho)\,\Delta\,\rho+{\widetilde f}(\rho)
\,\left({\bfm\nabla}\,\rho\right)^2 \right]\phi+V({\bfm x})\,\phi
\ , \label{krDGd}
\end{eqnarray}
with
\begin{equation}
{\widetilde f}(\rho)=b_+\,
\rho^{r+\kappa-1}-b_-\,\rho^{r-\kappa-1} \ ,
\end{equation}
and $b_\pm=a_\pm\,(r\pm\kappa-1/2)$.\\ Eq. (\ref{krDGd})
contains only a purely real nonlinearity and reduces to Eq.
(\ref{DG1}) in the $(\kappa,r)\to(0,\,0)$ limit, as well as
Eq. (\ref{skr}), which reduces to the
standard BG-entropy.\\
In particular, for Tsallis-entropy, the continuity equation
(\ref{r1}), with
\begin{equation}
f(\rho)=q\,\rho^{q-1} \ ,\label{Tsallis1}
\end{equation}
becomes the diffusive NFPE \cite{Compte1} while the
corresponding NSE is given through Eq. (\ref{krDGd}) with
\begin{eqnarray}
{\widetilde f}(\rho)=\left(q-{3\over2}\right)\,\rho^{q-2} \ ,
\end{eqnarray}
and reduces to Eq. (\ref{DG1}) in the $q\to1$ limit just as
entropy (\ref{tsallis}) reduces to BG-entropy.\\ We observe
that in Refs. \cite{Olavo1,Olavo} the quantization of a
classical system described by Tsallis-entropy has been
already discussed. There, a NLS compatible with the
continuity equation
$\partial\,\rho^\mu/\partial\,t+{\bfm\nabla}\cdot(\rho^\mu\,\widehat{\bfm
u}_{_{\rm drift}})=0$ was obtained with a different
approach. The nonlinearity appearing in the NLS of Refs.
\cite{Olavo1,Olavo} reduces,
for $\mu=1$ and $q\rightarrow2-q$, to the same one reported here.\\
On the other hand, for Kaniadakis-entropy, the continuity
equation is given in Eq. (\ref{r1}) with
\begin{equation}
f(\rho)={1\over2}\,\left[(\kappa+1)\,\rho^\kappa-(\kappa-1)\,\rho^{-\kappa}\right]
\ ,
\end{equation}
which coincides with that proposed in Ref.
\cite{Kaniadakis1} while the associated NSE is given in Eq.
(\ref{krDGd}) with
\begin{eqnarray}
{\widetilde
f}(\rho)={1\over2\,\rho}\,\left[(\kappa+1)\,\left(\kappa-{1\over2}\right)\,\rho^{\kappa}+
(\kappa-1)\,\left(\kappa+{1\over2}\right)\,\rho^{-\kappa}\right] \
,
\end{eqnarray}
and reduces to Eq. (\ref{DG1}) in the $\kappa\to0$ limit
just as entropy (\ref{sk}) reduces to BG-entropy.

\subsection{Interpolating bosons-fermions-entropy}
In Ref. \cite{Quarati1}, on the basis of the generalized
exclusion-inclusion principle the authors introduced a
family of NFPEs describing the evolution of a classical
system of particles whose statistical behavior interpolates
between bosonic and fermionic particles. The equilibrium
distribution governed by the EIP can be obtained by
maximizing the following entropy
\begin{equation}
S_{\rm EIP}(\rho)=-\int\left[\rho\,\ln\rho-
{1\over\kappa}\,(1+\kappa\,\rho)\,\ln(1+\kappa\,\rho)\right]\,d{\bfm
x} \ ,\label{seip}
\end{equation}
with $-1\leq\kappa\leq1$. In particular, for $\kappa=\pm1$ we
recognize the well-known Bose-Einstein and Fermi-Dirac entropies,
whereas intermediary behavior follows for $-1<\kappa<1$. Entropy
(\ref{seip}) can be obtained from Eq. (\ref{entropy}) by posing
$a(\rho)=\rho$ and $b(\rho)=1+\kappa\,\rho$.

Some examples of real physical systems where EIP can be
usefully applied are to be found in the Bose-Einsten
condensation. Typically, the cubic NSE is used to describe
the behavior of the condensate by simulating in this way
the statistical attraction between the many bodies
constituting the system. In spite of the simplest cubic
interaction, other interactions like the one introduced by
the EIP
can be adopted to simulate an attraction among the particles.\\
In the opposite direction, almost-fermionic systems can be
found in the study of the motion of electrons and holes in
a semiconductor. In fact, while if separately considered
electrons and holes are fermions, together they constitute
an excited state behaving differently from a fermion or a
boson. The same argument can be applied to the Cooper-pair
in the superconductivity theory. Such excitation differs
from a pure boson state because of the spatial
delocalization of the two electrons, which are not
completely overlying. Deviation from Bose statistics must
be taken into account.

In the following we discuss separately two different
choices for
functional $\gamma(\rho)$.\\
In the linear drift case, with
$c(\rho)=1/(1+\kappa\,\rho)$, the evolution equation for
field $\rho$ assumes the expression
\begin{equation}
\frac{\partial \rho}{\partial t}+{\bfm\nabla}\cdot\left({\bfm
j}_{_0}-D\,{{\bfm\nabla}\,\rho\over1+\kappa\,\rho}\,\right)=0 \
,\label{eip1}
\end{equation}
which was proposed in Ref. \cite{Kaniadakis}. By means of
Eq. (\ref{gauge}), nonlinear current ${\bfm
j}_{_0}-D\,{{\bfm\nabla}\,\rho/(1+\kappa\,\rho)}\rightarrow\widetilde{\bfm
j}_{_0}$ assumes the standard bilinear form and the
corresponding NSE follows from Eq. (\ref{schroedinger3})
\begin{eqnarray}
i\,\hbar\,\frac{\partial\,\phi}{\partial\,t}=&-&\frac{\hbar^2}{2\,m}\,\Delta\,\phi+
\frac{m\,D^2}{(1+\kappa\,\rho)^2}\,\left[
\frac{\Delta\,\rho}{\rho}-\frac{1-3\,\kappa\,\rho}{2\,(1+\kappa\,\rho)}\,\left(\frac{{\bfm\nabla}
\,\rho}{\rho}\right)^2 \right]\,\phi+V({\bfm x})\,\phi \ .
\label{qeip1}
\end{eqnarray}
We can observe that in Eq. (\ref{qeip1}) the EIP is
accounting through a diffusion process and its effect
vanishes in the $D\to0$ limit where it reduces to the
standard linear Schr\"odinger equation. Eq. (\ref{qeip1})
has a purely real nonlinearity depending only on field
$\rho$.

In a different way, by making the choice $c(\rho)=1$, the
continuity equation (\ref{r1}) becomes
\begin{equation}
\frac{\partial \rho}{\partial t}+{\bfm\nabla}\cdot\Big[{\bfm
j}_{_0}\,(1+\kappa\,\rho)-D\,{\bfm\nabla}\,\rho\Big]=0 \
.\label{eip2}
\end{equation}
The gauge transformation changes nonlinear current ${\bfm
j}\equiv{\bfm
j}_{_0}\,(1+\kappa\,\rho)-D\,{\bfm\nabla}\,\rho\rightarrow\widetilde{\bfm
j}\equiv\widetilde{\bfm j}_{_0}\,(1+\kappa\,\rho)$
containing only a nonlinear drift term and Eq. (\ref{eip2})
reduces to
\begin{equation}
\frac{\partial \rho}{\partial
t}+{\bfm\nabla}\cdot\left[\widetilde{\bfm
j}_{_0}\,(1+\kappa\,\rho)\right]=0 \ .\label{eip4}
\end{equation}
This equation was introduced at the classical level in Ref.
\cite{Quarati1} and subsequently reconsidered at the
quantum level in Ref. \cite{Scarfone3}. The NSE associated
with Eq. (\ref{eip4}) is given by
\begin{eqnarray}
\nonumber
i\,\hbar\,\frac{\partial\,\phi}{\partial\,t}=&-&\frac{\hbar^2}{2\,m}\,\Delta\,\phi
+\frac{m\,D^2}{1+\kappa\,\rho}\left[\frac{\Delta\,\rho}{\rho} -
\frac{1+2\,\kappa\,\rho}{2\,(1+\kappa\,\rho)}\,\left(\frac{{\bfm\nabla}
\,\rho}{\rho}\right)^2 \right]\,\phi\\
&+&\kappa\,{m\over\rho}\,\left(\frac{\widetilde{\bfm
j}}{1+\kappa\,\rho}\right)^2\,\phi-i\,\kappa\,
\frac{\hbar}{2\,\rho}\,{\bfm\nabla}\cdot\left(\frac{\widetilde{\bfm
j}\,\rho}{1+\kappa\,\rho} \right)\,\phi+V({\bfm x})\,\phi \
.\label{eip3}
\end{eqnarray}
We observe that Eq. (\ref{eip3}) still has a complex
nonlinearity due to the nonlinear structure of quantum
current $\widetilde{\bfm j}$ and both the nonlinearities
$W$ and $\cal W$ depend on fields $\rho$ and $\sigma$.
Moreover, in Eq. (\ref{eip3}), EIP is accounted through a
nonlinear drift term and survives even in absence of a
diffusion process $(D\to0)$.\\Factor $(1+\kappa\,\rho)$ in
nonlinear current $\widetilde{\bfm j}$ takes into account
the EIP in the many particle system. In fact, transition
probability (\ref{tran}) from site $\bfm x$ to $\bfm y$ is
defined as $\pi(t,\,{\bfm x}\rightarrow{\bfm
y})=r(t,\,{\bfm x},\,{\bfm x}\rightarrow{\bfm
y})\,\rho(t,\,{\bfm x})\,[1+\kappa\,\rho(t,\,{\bfm y})]$.
For $\kappa\not=0$ the EIP holds and parameter $\kappa$
quantifies to what extent particle
kinetics is affected by the particle population of the arrival site.\\
If $\kappa>0$ the $\pi(t,\,{\bfm x}\rightarrow{\bfm y})$
contains an inclusion principle. In fact, the population
density at arrival point ${\bfm y}$ stimulates the particle
transition and therefore transition probability increases
linearly with $\rho(t,\,{\bfm y})$. Where $\kappa<0$ the
$\pi(t,\,{\bfm x}\rightarrow{\bfm y})$ takes into account
the Pauli exclusion principle.  If the arrival point ${\bfm
y}$ is empty $\rho(t,\,{\bfm y})=0$, the $\pi(t,\,{\bfm
x}\rightarrow{\bfm y})$ depends only on the population of
the starting point. If arrival site is populated
$0<\rho(t,\,{\bfm y})\leq\rho_{max}$, the transition is
inhibited. The range of values that parameter $\kappa$ can
assume is limited by the condition that $\pi(t,\,{\bfm
x}\rightarrow{\bfm y})$ be real and positive as
$r(t,\,{\bfm x},\,{\bfm x}\rightarrow{\bfm y})$.  We may
conclude that $\kappa\geq-1/\rho_{max}$.

A physical meaning of parameter $\kappa$ can be supplied by
the following considerations. We recall that Bose-Einstein
and Fermi-Dirac statistics originate from the fundamental
principle of indistinguishability in quantum mechanics
which is closely related to the symmetrization of the wave
function. Completely symmetric wave functions are used to
describe bosons while fermions are described by completely
anti-symmetric wave functions. Thus, intermediate
statistics arise in presence of incomplete symmetrization
or anti-symmetrization of the wave function and the concept
of degree of symmetrization or degree of
anti-symmetrization has been introduced \cite{Quarati1}.
Parameter $\kappa$ has the meaning of degree of
indistinguishability of fermions or bosons, corresponding
to the degree of symmetrization or anti-symmetrization,
respectively. Value $\kappa=-1$ corresponds to the case of
fermions and denotes a complete anti-symmetric wave
function whereas value $\kappa=1$ corresponds to the case
of bosons and denotes a complete symmetric wave function.
In addition, value $\kappa=0$ is associated with classical
MB statistics and all the intermediate cases arise when
$\kappa$ assumes all the values between $-1$ and $1$.

Eq. (\ref{eip3}), for $D=0$, was obtained previously in
Ref. \cite{Scarfone3}, where the canonical quantization of
the classical system obeying EIP was accounted for. As
discussed in Section \ref{Gauge equivalence}, Eq.
(\ref{eip3}) differs from the NSE obtained in
\cite{Scarfone3} for a real nonlinearity originated from
nonlinear potential ${\widetilde
U}(\rho)=-m\,D^2\,({\bfm\nabla}\,\rho)^2/\rho\,(1+\kappa\rho)$
and depending only on field $\rho$.\\ Finally, we observe
that different from Eq. (\ref{qeip1}), Eq. (\ref{eip3}) has
vorticity different from zero. The Clebsh potentials
corresponding to current $\widetilde{\bfm
j}=({\bfm\nabla}\sigma/m)\,\rho\,(1+\kappa\,\rho)$ are
given by $\nu=1+\kappa\,\rho$, $\lambda=\sigma$ and
$\mu=const$ and vorticity assumes the expression
\begin{equation}
{\bfm\omega}=\frac{\kappa}{m}\,{\bfm\nabla}\,\rho\times{\bfm\nabla}\,\sigma
\ .\label{vorteip}
\end{equation}
In Ref. \cite{Scarfone5,Scarfone6} localized, static,
fermion-like vortex solutions ($\kappa<0$) were obtained
and studied starting from Eq. (\ref{eip3}) with $D=0$. We
observe that in \cite{Scarfone5,Scarfone6} a different
definition of the Clebsh potentials corresponding to
$\mu=\lambda=\sigma$ and $\nu=\kappa\,\rho$ was adopted.
Despite this, vorticity assumes the
same expression that is given by Eq. (\ref{vorteip}) in
both cases.\\
EIP vortex solutions are important on the theoretical
ground and for interpretation of experimental results of
several applications. For instance, they can be employed in
the study of fermion-like vortices observed in $^3$He-A
superfluidity or in heavy fermion superconductors UPt$_3$
and U$_{0.97}$Th$_{0.03}$Be$_{13}$
\cite{Williams,Matthews,Madison}.

\sect{Conclusions} We have presented the quantization of a
classical system of interacting particles obeying a kinetic
interaction principle. The KIP both fixes the expression of
the Fokker-Planck equation describing the kinetic evolution
of the system and imposes
the form of its entropy.\\
In the framework of canonical quantization, we have
introduced a class of NSEs with complex nonlinearity
obtained from the classical system obeying KIP. The form of
nonlinearity $\Lambda(\psi^\ast,\,\psi)$ is determined by
functional $\kappa(\rho)$, which also fixes the form of the
entropy of ancestor classical system.

Among the many interesting solutions of the family of NSEs
(\ref{schroedinger2}) we observe that for a free system
with $V({\bfm x})=0$, and posing $G(\rho)=0$, the planar
wave
\begin{equation}
\psi(t,\,{\bfm x})=A\,\exp\left(-\frac{i}{\hbar}\,(\omega\,t-{\bfm
k}\cdot{\bfm x})\right) \ ,
\end{equation}
with constant amplitude $A=const$ is the simplest solution, where
the relationship between $\omega$ and $\bfm k$ is given by
\begin{equation}
\omega=\frac{\hbar^2\,{\bfm
k}^2}{2\,m}\,\frac{\partial\,\gamma(\rho)}{\partial\,\rho}\Bigg|_{\rho=A^2}
\ ,
\end{equation}
and reduces to the standard dispersion relation for
$\gamma(\rho)=\rho$.

When the quantum system is in a stationary state such that
$\partial\,\rho_{\rm s}/\partial\,t=0$, the relationships
between distribution $\rho_{\rm s}$ and phase $\Sigma_{\rm
s}$ follow from Eq. (\ref{r1})
\begin{equation}
\rho_{_{\rm s}}=\kappa^{-1}\left(\exp\left(\frac{\Sigma_{\rm
s}({\bfm x})}{m\,D}-\beta^\prime\right)\right) \ ,\label{equi1}
\end{equation}
which mimics the classical equilibrium distribution
(\ref{equi}), as can be seen by replacing $\Sigma_{\rm
s}({\bfm x})/m\,D$ with $-\beta\,{\cal E}({\bfm x})$.
Despite this, we stress that such an analogy is purely
formal. The equivalence between Eqs. (\ref{equi}) and
(\ref{equi1}) requires that the following relation
$\Sigma_{\rm s}({\bfm x})/m\,D=-\beta\,{\cal E}({\bfm x})$
must hold. In the general case the expression of stationary
phase $\Sigma_{\rm s}({\bfm x})$ must be obtained from Eq.
(\ref{s1}), after posing $\partial\,\Sigma_{\rm
s}/\partial\,t=0$, with $\rho$ given through Eq.
(\ref{equi1}).

Finally, another interesting class of possible solutions
are solitons. It is well know that soliton solutions in NSE
arise when the dispersive effects, principally due to term
$-(\hbar^2/2\,m)\,\Delta\,\psi$, is exactly balanced by the
nonlinear terms. The existence of this class of solutions
depends on the particular form of functionals
$\gamma(\rho)$ and $\kappa(\rho)$ which fix the expression
of nonlinearities $W(\rho,\,\Sigma)$ and ${\cal
W}(\rho,\,\Sigma)$. A special situation, where soliton
solutions are found within the NSEs derived in this paper,
is given by the EIP-equation (\ref{eip3}) with $D=0$
\cite{Scarfone3}
where $\gamma(\rho)=\rho\,(1+\kappa\,\rho)$ and $\kappa(\rho)=\rho/(1+\kappa\,\rho)$.\\
The study of soliton solutions for other functional choices
of $\gamma(\rho)$ and $\kappa(\rho)$, like, for instance,
the ones related to the generalized entropies discussed in
Section VI-B, is a very important task which deserves
further research. These solutions may lead to practical
applications. In fact, in recent years there has been great
interest in the formulation of models where solitons can
interact with a long-range force \cite{Gaididei}. Typical
nonlinear models supporting solitons, like the sine-Gordon
model, arise from short-range forces. However, there is
experimental evidence that most real transfer mechanisms
have long-range interaction, as noted in condensed matter
theory \cite{Scott} or in spin glasses \cite{Ford}.

\app \sect{}

We present proof of the Ehrenfest equations discussed in
Section \ref{Ehrenfest}. In the following we assume uniform
boundary conditions on the fields in order to neglect the
surface terms.

Let us rewrite Eq. (\ref{ehrenfest}) in a more suitable form.
Accounting for the relation
\begin{equation}
\frac{\delta}{\delta\,\psi}=\psi^\ast\left(\frac{\delta}{\delta\,\rho}
-\frac{i\,\hbar}{2\,\rho}\,\frac{\delta}{\delta\,\Sigma}\right) \
,
\end{equation}
Eq. (\ref{ehrenfest}) becomes
\begin{equation}
\frac{d}{d\,t}\langle{\cal
O}\,\rangle={i\over\hbar}\Bigg\langle\left[\frac{\delta\,H}{\delta\,\rho},\,{\cal
O}\right]\Bigg\rangle+{1\over2}\,\Bigg\langle\left\{{1\over\rho}\,
\frac{\delta\,H}{\delta\,\Sigma},\,{\cal
O}\right\}\Bigg\rangle+\Bigg\langle\frac{\partial\,{\cal
O}}{\partial\,t}\Bigg\rangle \ .\label{ehrenfest1}
\end{equation}

Eq. (\ref{prima}) can be obtained starting from Eq.
(\ref{ehrenfest1}) by posing ${\cal O}={\bfm x}$
\begin{eqnarray}
\nonumber \frac{d}{d\,t}\,\langle{\bfm
x}\rangle&=&{i\over\hbar}\int\left[\psi^\ast\,
\frac{\delta\,H}{\delta\,\rho}\,{\bfm x}\,\psi-\psi^\ast\,{\bfm
x}\,\frac{\delta\,H}{\delta\,\rho}\,\psi\right]\,d{\bfm
x}+\frac{1}{2}\int\left[\psi^\ast\,{1\over\rho}\,
\frac{\delta\,H}{\delta\,\Sigma}\,{\bfm x}\,\psi+\psi^\ast\,{\bfm
x}\,{1\over\rho}\,\frac{\delta\,H}{\delta\,\Sigma}\,\psi\right]\,d{\bfm
x}\\
\nonumber&=&\int {\bfm
x}\,\frac{\delta\,H}{\delta\,\Sigma}\,d{\bfm
x}\\
\nonumber &=&-\int{\bfm x}\,{\bfm\nabla}\cdot \left[\frac{{\bfm
\nabla}\,\Sigma}{m}\,\gamma(\rho)-D\,f(\rho)
\,{\bfm\nabla}\,\rho\right]\,d{\bfm x}\\
\nonumber &=&\int\left[\frac{{\bfm
\nabla}\,\Sigma}{m}\,\gamma(\rho)-D\,f(\rho)
\,{\bfm\nabla}\,\rho\right]\,d{\bfm x}\\
\nonumber &=&\int\frac{{\bfm
\nabla}\,\Sigma}{m}\,\gamma(\rho)\,d{\bfm x}-D\int
{\bfm\nabla}\,F(\rho)\,d{\bfm x}\\
 &=&\Bigg\langle \frac{\gamma(\rho)}{\rho}\,\widehat{\bfm
u}_{\rm drift}\Bigg\rangle \ ,
\end{eqnarray}
where
\begin{equation}
F(\rho)=\int^\rho f(\rho^\prime)\,d\rho^\prime \ .
\end{equation}
 To show the validity of Eq.
(\ref{seconda}) we pose ${\cal O}=-i\,\hbar\,{\bfm\nabla}$ in Eq.
(\ref{ehrenfest1}) so that
\begin{eqnarray}
\nonumber \frac{d}{d\,t}\,\langle{\bfm
p}\rangle&=&\int\left[\psi^\ast\,
\frac{\delta\,H}{\delta\,\rho}\,{\bfm\nabla}\,\psi-\psi^\ast\,{\bfm
\nabla}\left(\frac{\delta\,H}{\delta\,\rho}\,\psi\right)\right]\,d{\bfm
x}\\
\nonumber &-&i\,\frac{\hbar}{2}\int\left[\psi^\ast\,{1\over\rho}\,
\frac{\delta\,H}{\delta\,\Sigma}\,{\bfm\nabla}\,\psi+\psi^\ast\,{\bfm
\nabla}\left({1\over\rho}\,\frac{\delta\,H}{\delta\,\Sigma}\,\psi\right)\right]\,d{\bfm
x}\\
\nonumber &=&\int
\frac{\delta\,H}{\delta\,\rho}\,\left(\psi^\ast\,{\bfm\nabla}\,\psi+
\psi\,{\bfm\nabla}\,\psi^\ast\right)-i\,\frac{\hbar}{2}\int
{1\over\rho}\,\frac{\delta\,H}{\delta\,\Sigma}\,\left(\psi^\ast\,{\bfm\nabla}\,\psi-
\psi\,{\bfm\nabla}\,\psi^\ast\right)\,d{\bfm
x}\\
&=&\int\left(\frac{\delta\,H}{\delta\,\rho}\,{\bfm\nabla}\,\rho+
\frac{\delta\,H}{\delta\,\Sigma}\,{\bfm\nabla}\,\Sigma\right)\,d{\bfm
x} \ ,\label{a2}
\end{eqnarray}
where an integration by parts has been performed, and we have
posed
\begin{eqnarray}
&&\psi^\ast\,{\bfm\nabla}\,\psi+\psi\,{\bfm\nabla}\,\psi^\ast
={\bfm\nabla}\,\rho \ ,\\
&&\psi^\ast\,{\bfm\nabla}\,\psi-\psi\,{\bfm\nabla}\,\psi^\ast
=i\,\frac{2}{\hbar}\,\rho\,{\bfm\nabla}\,\Sigma \ .
\end{eqnarray}
Taking into account the relation
\begin{eqnarray}
{\bfm\nabla}\,{\cal
H}=\frac{\delta\,H}{\delta\,\rho}\,{\bfm\nabla}\,\rho+
\frac{\delta\,H}{\delta\,\Sigma}\,{\bfm\nabla}\,\Sigma+\rho\,{\bfm\nabla}\,V({\bfm
x}) \ ,
\end{eqnarray}
from Eq. (\ref{a2}) it follows
\begin{eqnarray}
\nonumber \frac{d}{d\,t}\,\langle{\bfm
p}\rangle&=&\int{\bfm\nabla}\,{\cal H}\,d{\bfm
x}-\int\rho\,{\bfm\nabla}\,V({\bfm x})\,d{\bfm x}\\&=&\langle{\bfm
F}_{\rm ext}({\bfm x})\rangle \ ,\label{sec}
\end{eqnarray}

Eq. (\ref{terza}) can easily be obtained following the same
steps used in the proof of Eq. (\ref{seconda}).

Finally, by posing ${\cal
O}=-(\hbar^2/2\,m)\,\Delta+U(\rho,\,\Sigma)/\rho+V({\bfm
x})$ in Eq. (\ref{ehrenfest1}), where $U(\rho,\,\Sigma)$ is
given in Eq. (\ref{u}), we have
\begin{eqnarray}
\nonumber
\frac{d\,E}{d\,t}&=&{i\over\hbar}\,\int\left\{\psi^\ast\,\frac{\delta\,H}{\delta\,\rho}\,
\left(-\frac{\hbar^2}{2\,m}\,\Delta+{U\over\rho}+V\right)\,\psi
-\psi^\ast\,\left[\left(-\frac{\hbar^2}{2\,m}\,\Delta+
{U\over\rho}+V\right)\,\frac{\delta\,H}{\delta\,\rho}\,\psi\right]\right\}\,d{\bfm x}\\
\nonumber &+
&{1\over2}\int\left\{\psi^\ast\,{1\over\rho}\,\frac{\delta\,H}{\delta\,\Sigma}\,
\left(-\frac{\hbar^2}{2\,m}\,\Delta+{U\over\rho}+V\right)\,\psi
+\psi^\ast\,\left[\left(-\frac{\hbar^2}{2\,m}\,\Delta+
{U\over\rho}+V\right)\,{1\over\rho}\,
\frac{\delta\,H}{\delta\,\rho}\,\psi\right]\right\}\,d{\bfm x}\\
\nonumber &+
&\int\rho\,\frac{\partial}{\partial\,t}\left(\frac{U}{\rho}+V\right)\,d{\bfm
x}\\
\nonumber
&=&-\frac{i\,\hbar}{2\,m}\,\int\left[\psi^\ast\,\frac{\delta\,H}{\delta\,\rho}\,
\Delta\,\psi-\psi^\ast\,\Delta\left(\frac{\delta\,H}{\delta\,\rho}\,\psi\right)\right]\,d{\bfm
x}\\
\nonumber &+
&{i\over\hbar}\int\left[\psi^\ast\,\frac{\delta\,H}{\delta\,\rho}\,
\left({U\over\rho}+V\right)\,\psi
-\psi^\ast\,\left({U\over\rho}+V\right)\,\frac{\delta\,H}{\delta\,\rho}\,\psi\right]\,d{\bfm
x}\\ \nonumber
&-&\frac{\hbar^2}{4\,m}\int\left[\psi^\ast\,{1\over\rho}\,\frac{\delta\,H}{\delta\,\Sigma}\,
\Delta\,\psi+\psi^\ast\,\Delta\left({1\over\rho}\,\frac{\delta\,H}{\delta\,\Sigma}\,\psi\right)
\right]\,d{\bfm
x}\\
\nonumber&+&{1\over2}\int\left[\psi^\ast\,{1\over\rho}\,\frac{\delta\,H}{\delta\,\Sigma}\,
\left({U\over\rho}+V\right)\,\psi
+\psi^\ast\,\left({U\over\rho}+V\right)\,{1\over\rho}\,\frac{\delta\,H}{\delta\,\Sigma}
\,\psi\right]\,d{\bfm
x}+\int\rho\,\frac{\partial}{\partial\,t}\left({U\over\rho}\right)\,d{\bfm
x}\\
\nonumber&=&-\frac{i\,\hbar}{2\,m}\,\int\frac{\delta\,H}{\delta\,\rho}\,\left(\psi^\ast\,
\Delta\,\psi-\psi\,\Delta\,\psi^\ast\right)\,d{\bfm
x}-\frac{\hbar^2}{4\,m}\int{1\over\rho}\,\frac{\delta\,H}{\delta\,\Sigma}\,
\left(\psi^\ast\,
\Delta\,\psi+\psi\,\Delta\,\psi^\ast\right)\,d{\bfm
x}\\
&+&\int\left[\frac{\delta\,H}{\delta\,\Sigma}\,\left({U\over\rho}+V\right)+
\frac{\partial\,U}{\partial\,t}-{U\over\rho}\,\frac{\partial\,\rho}{\partial\,t}\right]\,d{\bfm
x} \ ,\label{a4}
\end{eqnarray}
where a double integration by parts has been performed.
Taking into account
\begin{eqnarray}
&&-\frac{i\,\hbar}{2\,m}\,\left(\psi^\ast\,\Delta\,\psi-\psi\,\Delta\,\psi^\ast\right)=
{\bfm\nabla}\cdot\left(\frac{{\bfm\nabla}\,\Sigma}{m}\,\rho\right) \ ,\\
&&\psi^\ast\,\Delta\,\psi+\psi\,\Delta\,\psi^\ast=2\,\rho\,\left[\frac{\Delta\,\sqrt{\rho}}
{\sqrt{\rho}}-\left(\frac{{\bfm\nabla}\,\Sigma}{\hbar}\right)^2\right]
\ ,
\end{eqnarray}
which follow from Eq. (\ref{polar}), and the relation
\begin{equation}
\frac{\partial\,U}{\partial\,t}=\left(\frac{\delta}{\delta\,\rho}\int\,U\,d{\bfm
x}\right)\,
\frac{\partial\,\rho}{\partial\,t}+\left(\frac{\delta}{\delta\,\Sigma}\int\,U\,d{\bfm
x}\right)\, \frac{\partial\,\Sigma}{\partial\,t} \ ,
\end{equation}
Eq. (\ref{a4}) becomes
\begin{eqnarray}
\nonumber
\frac{d\,E}{d\,t}&=&\int\left\{\frac{\delta\,H}{\delta\,\rho}\,
{\bfm\nabla}\cdot\left(\frac{{\bfm\nabla}\,\Sigma}{m}\,\rho\right)-
\frac{\hbar^2}{2\,m}\,\frac{\delta\,H}{\delta\,\Sigma}\,\left[\frac{\Delta\,\sqrt{\rho}}
{\sqrt{\rho}}-\left(\frac{{\bfm\nabla}\,\Sigma}{\hbar}\right)^2\right]\right\}\,d{\bfm
x}\\
\nonumber
&+&\int\left[\frac{\delta\,H}{\delta\,\Sigma}\,\left({U\over\rho}+V\right)+
\left(\frac{\delta}{\delta\,\rho}\int\,U\,d{\bfm
x}-{U\over\rho}\right)\,\frac{\partial\,\rho}{\partial\,t}
+\left(\frac{\delta}{\delta\,\Sigma}\int\,U\,d{\bfm x}\right)\,
\frac{\partial\,\Sigma}{\partial\,t}\right]\,d{\bfm x} \
.\\\label{a5}
\end{eqnarray}
By using the relations
\begin{eqnarray}
&&\frac{\hbar^2}{2\,m}\,\left[\frac{\Delta\,\sqrt{\rho}}{\sqrt{\rho}}-
\left(\frac{{\bfm\nabla}\,\Sigma}{\hbar}\right)^2
\right]=\frac{\delta}{\delta\,\rho}\int\,U\,d{\bfm x}-
\frac{\delta\,H}{\delta\rho}+V \ ,\\
&&{\bfm\nabla}\cdot\left(\frac{{\bfm\nabla}\,\Sigma}{m}\,\rho\right)=
\frac{\delta}{\delta\,\Sigma}\int\,U\,d{\bfm x}-
\frac{\delta\,H}{\delta\Sigma} \ ,
\end{eqnarray}
which follow from Eqs. (\ref{ha}), (\ref{ham1}) and
(\ref{u}), and motion equations (\ref{rhos1}) and
(\ref{rhos2}), we obtain
\begin{eqnarray}
\nonumber
\frac{d\,E}{d\,t}&=&\int\left[\frac{\delta\,H}{\delta\,\rho}\,
\left(\frac{\delta}{\delta\,\Sigma}\int\,U\,d{\bfm x}-
\frac{\delta\,H}{\delta\Sigma}\right)-
\frac{\delta\,H}{\delta\,\Sigma}\,\left(\frac{\delta}{\delta\,\rho}\int\,U\,d{\bfm
x}- \frac{\delta\,H}{\delta\rho}+V\right)\right]\,d{\bfm
x}\\
\nonumber
&+&\int\left[\frac{\delta\,H}{\delta\,\Sigma}\,\left({U\over\rho}+V\right)+
\frac{\delta\,H}{\delta\Sigma}\,\left(\frac{\delta}{\delta\,\rho}\int\,U\,d{\bfm
x}-{U\over\rho}\right) -
\frac{\delta\,H}{\delta\rho}\,\left(\frac{\delta}{\delta\,\Sigma}\int\,U\,d{\bfm
x}\right)\right]\,d{\bfm
x}\\
&=&0 \ .
\end{eqnarray}


\sect{}

We briefly discuss the generalization of the theory for
quantum systems obeying the KIP and undergoing a diffusive
process with a diffusion coefficient $D(t,\,{\bfm
x})$ depending both on time and space position.\\
Given the following Hamiltonian density
\begin{eqnarray}
\nonumber {\cal
H}(\rho,\,\Sigma)=\frac{({\bfm\nabla}\,\Sigma)^2}{2\,m}\,\gamma(\rho)+\frac{\hbar^2}{8\,m}
\,\frac{\left({\bfm\nabla}\,\rho\right)^2}{\rho}-D(t,\,{\bfm
x})\,\gamma(\rho)\,{\bfm\nabla}\,\ln\kappa(\rho)\cdot{\bfm\nabla
\Sigma}+{\widetilde U}(\rho)+V({\bfm x})\,\rho \ ,\\ \label{hh}
\end{eqnarray}
from the Hamilton equations (\ref{rhos1})-(\ref{rhos2}) we obtain
the NSE
\begin{equation}
i\,\hbar\,\frac{\partial\,\psi}{\partial\,t}=-\frac{\hbar^2}{2\,m}\,\Delta\,\psi+
\Big[W(\rho,\,\Sigma)\,+i\,{\cal
W}(\rho,\,\Sigma)\Big]\,\psi+G(\rho)\,\psi+V({\bfm x})\,\Psi \ ,
\end{equation}
with nonlinearities
\begin{eqnarray}
\hspace{-3mm}W(\rho,\,\Sigma)={m\over2}\,
\left(\frac{\partial\,\gamma(\rho)}{\partial\,\rho}-1\right)\,\left(\frac{{\bfm
j}_0}{\rho}\right)^2+m\,\gamma(\rho)\,\frac{\partial}{\partial\,\rho}\,\ln\kappa(\rho)\,
{\bfm\nabla}\cdot\left(D(t,\,{\bfm x})\,\frac{{\bfm
j}_{_0}}{\rho}\right)+G(\rho) \ ,\label{b1}
\end{eqnarray}
and
\begin{eqnarray}
{\mathcal
W}(\rho,\,\Sigma)=-\frac{\hbar}{2\,m\,\rho}\,{\bfm\nabla}\,\Big\{[\gamma(\rho)-\rho]
\,{\bfm\nabla}\,\Sigma\Big\}
+\frac{1}{2\,\rho}\,{\bfm\nabla}\cdot\left[D(t,\,{\bfm
x})\,\gamma(\rho ) \,{\bfm\nabla}\,\ln\kappa(\rho)\right] \
.\label{b2}
\end{eqnarray}
The system described by Hamiltonian (\ref{hh}) is
dissipative since $d\,E/d\,t\not=0$. This is a consequence
of the time dependence of $D$ which breaks the invariance
of Eq. (\ref{hh}) under uniform time translation. In the
same way, linear momentum as well as angular momentum are
no longer conserved, even in absence of the external
potential, as a consequence of the position dependence of
$D$ which breaks the invariance of Eq. (\ref{hh}) under
uniform space translation and uniform space rotation.  This
can also be seen from the Ehrenfest relations
\begin{eqnarray}
&&\frac{d}{d\,t}\,\langle{\bfm x}\rangle=\Bigg\langle
\frac{\gamma(\rho)}{\rho}\,\widehat{\bfm u}_{\rm
drift}\Bigg\rangle-\Bigg\langle D(t,\,{\bfm
x})\,f(\rho)\,{\bfm\nabla}\,\ln\rho\Bigg\rangle \
,\label{dprima}\\
&&\frac{d}{d\,t}\,\langle{\bfm p}\rangle=-m\,\Big\langle
A(\rho,\,\Sigma)\,{\bfm\nabla}\,D(t,\,{\bfm
x})\Big\rangle+\Big\langle{\bfm F}_{\rm
ext}({\bfm x})\Big\rangle \ ,\label{dseconda}\\
&&\frac{d}{d\,t}\,\langle{\bfm L}\rangle=-m\,\Big\langle
A(\rho,\,\Sigma)\,\Big({\bfm x}\times{\bfm\nabla}\,D(t,\,{\bfm
x})\Big)\Big\rangle+\Big\langle{\bfm M}_{\rm
ext}({\bfm x})\Big\rangle \ ,\label{dterza}\\
&&\frac{d\,E}{d\,t}=-m\,\Big\langle
A(\rho,\,\Sigma)\,{\partial\over\partial\,t}\,D(t,\,{\bfm
x})\Big\rangle \ ,\label{dquarta}
\end{eqnarray}
where $A(\rho,\,\Sigma)=f(\rho)\,{\bfm\nabla}\,\ln
\rho\cdot\,\widehat{\bfm u}_{\rm drift}$.

Finally, the gauge transformation described in Section V
cannot be performed, in general, when the diffusion
coefficient has spatial dependence. In fact, the
transformation in (\ref{gauge}) is well defined only if the
following condition is fulfilled
\begin{equation}
{\bfm\nabla}\times\Big[D(t,\,{\bfm
x})\,{\bfm\nabla}\,\ln\kappa(\rho)\Big]=0 \ ,\label{b9}
\end{equation}
as can be seen by applying the curl operator to both sides
of equation
\begin{equation}
{\bfm\nabla}\,\sigma={\bfm\nabla}\,\Sigma-m\,D(t,\,{\bfm
x})\,{\bfm\nabla}\,\ln\kappa(\rho) \ ,
\end{equation}
which follows from Eqs. (\ref{ncurrent}) and (\ref{jt}). We
remark that if the dynamics of the system evolves in one
spatial dimension, Eq. (\ref{b9}) is trivially verified and
the transformation in (\ref{gauge}) can in all cases be
accomplished.


\end{document}